\documentclass[onecolumn,12pt,journal,twoside,draftclsnofoot]{IEEEtranTCOM}
\usepackage{ifpdf}
\usepackage{cite}
\usepackage[pdftex]{graphicx}
\usepackage{algorithmic}
\usepackage{array}
\usepackage{stfloats}
\usepackage{color}
\usepackage{booktabs}
\usepackage{amsmath}
\usepackage{amsfonts}
\usepackage{amssymb}
\usepackage{amsthm}
\usepackage{tikz}
\usepackage{bm}
\usepackage{multirow}
\usepackage{makecell}
\usepackage{mathdots}
\usepackage{xtab}
\usepackage{caption}
\addtocounter{MaxMatrixCols}{10}
\theoremstyle{plain}

\newtheorem{theorem}{Theorem}
\newtheorem{lemma}[theorem]{Lemma}

\newtheorem{definition}[theorem]{Definition}
\newtheorem{corollary}[theorem]{Corollary}
\theoremstyle{definition}
\newtheorem{example}{Example}

\newtheorem{remark}{Remark}

\renewcommand{\tablename}{\scriptsize Table}
\renewcommand{\thetable}{\arabic{table}}
\begin{document}
\title{Folded Polynomial Codes for Coded Distributed $AA^\top$-Type  Matrix Multiplication} \author{
Jingke Xu,
Yaqian Zhang,
Libo Wang
\thanks{Jingke Xu is with School of Information Science and Engineering, Shandong Agricultural University, Taian 271018, China, e-mail: xujingke@sdau.edu.cn}
\thanks{Yaqian Zhang is with School of Electronic Information and Electrical Engineering, Shanghai Jiao Tong University, Shanghai 200240, China,  e-mail: zhangyq9@sjtu.edu.cn}
\thanks{Libo Wang is with College of Information Science and Technology, Jinan University, Guangzhou 510632, China, e-mail: wanglibo12b@mails.ucas.edu.cn}
}
\markboth{IEEE Transactions on Communications}%
{Submitted paper}
\maketitle

\begin{abstract}
In this paper, due to the important value in practical applications, we consider the coded distributed matrix multiplication problem of computing $AA^\top$ in a distributed computing system with $N$ worker nodes and a master node, where the input matrices $A$ and $A^\top$ are partitioned into $m$-by-$p$ and $p$-by-$m$ blocks of equal-size sub-matrices respectively. For effective straggler mitigation, we propose a novel computation strategy, named \emph{folded polynomial code}, which is obtained by modifying the entangled polynomial codes.
Moreover, we characterize a lower bound on the optimal recovery threshold among all linear computation strategies when the underlying field is the real number field, and our folded polynomial codes can achieve this bound in the case of $m=1$. Compared with all known computation strategies for coded distributed matrix multiplication, our folded polynomial codes outperform them in terms of recovery threshold, download cost, and decoding complexity.
\end{abstract}

 \IEEEpeerreviewmaketitle
\section{Introduction}
As the era of big data advances, coded distributed computing has emerged as an important approach to speed up large-scale data analysis tasks, such as machine learning, principal component analysis and graph processing. However, this may engender additional communication overhead, or lead to a computational bottleneck that arises due to the unpredictable latency in waiting for the slowest servers to finish their computations, referred to as straggler effect\cite{DB13,YHGR16}. It has been demonstrated in \cite{TLDK17},\cite{YHGR16} that stragglers may run  5 to 8 times slower than the typical worker on Amazon EC2 and thus cause significant delay in calculation. To mitigate straggler, researchers inject redundancy  by using error-correcting codes and design many coded computation strategies \cite{WJG15,DCG16},\cite{LLPPRIT18},\cite{LSR17}. The issue of straggling has become a hot topic in area of coded distribute computation \cite{DCG19,TLDK17,AAH19,APS17,LMA16}.

As a fundamental building block of many computing applications,  large scale distributed matrix multiplication has been used to process massive data in distributed computing frameworks, such as MapReduce\cite{DG04} and Apache Spark \cite{ZCFSS10}. The problem of coded distributed matrix multiplication (CDMM) can be formulated as  a user wants to compute the product $C=AB$ of two large data matrices $A$ and $B$ through a distributed computing system that consists of a master node and $N$ worker servers. The master node evenly divides $A$ and $B$ into block matrices and encodes them. Then, the master node assigns the encoded sub-matrices to the corresponding worker servers, who compute the product of encoded sub-matrices and return them to the master. After receiving the results from the worker servers, the master node can easily obtain the product $C$. To characterize the robustness against stragglers effects of a computation strategy, the recovery threshold \cite{YMA17} is defined as the minimum number of successful (non-delayed, non-faulty) worker servers that the master node needs to wait for completing the task.

Recently, straggler mitigation in CDMM has been deeply studied in the literature. The problem of CDMM was considered in  \cite{LLPPRIT18},\cite{LSR17} by using maximum distance separable codes to encode the input matrices. In \cite{YMA17}, Yu {\it et al.} designed a novel coded matrix-multiplication computation strategy, {\it Polynomial codes}, that outperformed classical works \cite{HA84,HR15} in algorithm-based fault tolerance (ABFT) with respect to  the recovery threshold. Polynomial codes have the recovery threshold $mn$, where the input matrix $A$ is vertically divided into $m$ row block sub-matrices and $B$ is horizontally divided into $n$ column block sub-matrices. While Dutta {\it et al.} in \cite{MatDotIT20} constructed  MatDot codes which reduced the recovery threshold to $2p-1$ for $p$ column-wise
partitioning of matrix $A$ and $p$ row-wise partitioning of $B$  at the cost of increasing the computational complexity and download  cost from each work server \cite{RDT20}. Entangled polynomial (EP) codes \cite{YMIT20} and generalized PolyDot codes\cite{DCG18} were independently constructed to build the general tradeoff between recovery threshold and download cost by arbitrarily partitioning the two data matrices into $m$-by-$p$ and $p$-by-$n$ blocks of equal-size sub-matrices respectively, which bridged the gap between Polynomial codes \cite{YMA17} and MatDot codes \cite{MatDotIT20}. Moreover, Yu {\it et al.} in \cite{YMIT20} also discussed the lower bound of the recovery threshold for linear coded strategies, and determined an entangled polynomial code achieving the optimal recovery threshold among all
linear coded strategies in the cases of $p=1$ or $m=n=1$. Furthermore, $\varepsilon$-approximate MatDot codes  in \cite{JDCP21} were designed  to compute $C=AB$ by allowing $\varepsilon$-error under the same partition of input matrices as in MatDot codes, which improved the recovery threshold from $2p-1$ to $p$. In addition to single computing task, there is also coded  distributed batch  matrix multiplication (CDBMM) \cite{JJ21},\cite{YAISIT20,YLR19},\cite{ZX21},\cite{TD22}.

As a special class of matrix multiplication,  $AA^\top$-type matrix multiplication is an important component in many scientific computations, such as image processing based on singular value decomposition \cite{S12}, and recommender systems based on the coded alternating least squares algorithm \cite{WYZWS21}. Actually,  the coded alternating least squares algorithm \cite{WYZWS21} repeatedly invoked $AA^\top$-type matrix multiplications, which were performed by using MatDot codes. However, to the best of our knowledge, the problem of coded  distributed $AA^\top$-type matrix multiplication has not been concerned and studied in depth so far. Therefore, no matter from practical applications or theoretical research, it is worthwhile to design efficient computation strategies with lower recovery threshold for such kind of matrix multiplication.
Actually, many large-scale scientific calculations are performed in the real number field, which implies that it is particularly important to characterize the optimal recovery threshold in this case. Thus, as a preliminary exploration in this area, we will investigate the optimal recovery threshold for all possible linear coded strategies over the real number field.

In this paper, we focus on the  coded distributed matrix multiplication problem of computing $AA^\top$ through a distributed computing system that consists of $N $ worker servers, where the input matrices $A$ and $A^\top$ over $\mathbb{F}$ are partitioned into $m$-by-$p$ and $p$-by-$m$ blocks of equal-size sub-matrices respectively\footnote{The aim  of such partition is to utilize the symmetry of the matrix $C=AA^\top$ in this paper.}. The main contributions of this paper are two folds:
 \begin{itemize}
 \item[1)] For arbitrary matrix partitioning with parameters $m,p$, a novel and efficient linear computation strategy, named\emph{ folded polynomial code},  is presented  by modifying the EP codes. The folded polynomial code achieves a recovery threshold $R_{\rm FP}= \binom{m+1}{2}+\frac{(p-1)(2m^2-m+1)+\chi(m)\chi(p)}{2}$, where $\chi(x)=\frac{1+(-1)^x}{2}$ for $x \in \mathbb{N}$.
 In particular, our folded polynomial code can be explicitly built for $m=1$ as long as $|\mathbb{F}|>2N$, and in this case its recovery threshold is $p$.
\item[2)] For all linear computation strategies of computing $AA^\top$-type matrix multiplication over the real number field $\mathbb{R}$, there is a lower bound on the recovery threshold $R$, i.e., $R\geq \min\{N, pm\}$. Particularly, our folded polynomial code  achieves the lower bound for $m=1$, namely, our lower bound is tight in this case.
\end{itemize}	

The rest of this paper is organized as follows. Section \ref{sec2} introduces the system model and some preliminaries. The main results are presented in Section III. In Section IV, the folded polynomial codes are constructed for general matrix partitioning. Comparisons between folded polynomial codes and previous schemes, and some numerical experiments are given in Section V. Finally, this paper is concluded in Section VI.

\section{Problem Statement and Preliminaries}\label{sec2}

\subsection{Notations and Problem Formulation}

Let $\mathbb{F}$ be a field and ${\rm char}\mathbb{F}$ be the characteristic of $\mathbb{F}$.  Denote the set of all polynomials with $n$ variables over $\mathbb{F}$  by $\mathbb{F}[x_1,x_2,\cdots,x_n]$. For $n_1<n_2\in \mathbb{N}$, denote $[n_1:n_2]=\{n_1,n_1+1,\cdots,n_2\}$. The cardinality of a set $\mathcal{S}$ is denoted by $|\mathcal{S}|$. We usually use capital letters to denote matrices (e.g., $A,B$) and bold lowercase letters to represent vectors (e.g., $\bm{a},\bm{b}$), and $A^{\top}$, $\bm{a}^{\top}$ denote the transpose of matrix $A$ and vector $\bm{a}$, respectively.
The $i$th element of a vector $\bm{a}$ is denoted by $a_i$ and $[A]_{i,j}$ represents the $(i,j)$th entry of a matrix $A$. For $\mathcal{X}=\{\beta_i\in\mathbb{F}:i\in[1:n]\}$ and $\mathcal{G}=\{g_i(x)\in\mathbb{F}[x]: i\in[1:k]\}$, define a matrix
${\rm M}(\mathcal{G},\mathcal{X})=(g_i(\beta_j))_{i\in[1:k], j\in[1:n]}\in\mathbb{F}^{k\times n}$ whose $(i,j)$-entry is $g_i(\beta_j)$.
Define ${\rm Span}(\mathcal{G})=\{\sum_{i=1}^ka_ig_i(x): a_i\in\mathbb{F},i\in[1:k]\}$, which is a linear space over $\mathbb{F}$.

We consider the matrix product $C=AA^\top$, which is computed in a distributed computing system consisting of a master node and $N$ worker nodes.
We assume that each worker node only connects to the master node and all the connected links are error-free and secure.

Formally, the master node first encodes its matrices for the $i$th worker node
as $\tilde{A}_i$ and $\tilde{B}_i$ according to the functions
$\bm{f}=(f_1,f_2,...,f_N)$ and ${\bm g}=(g_1,g_2,...,g_N)$, where
$\tilde{A}_i=f_i(A)$ and $\tilde{B}_i=g_i(A^\top)$ for $1\leq i \leq N$. After receiving the encoded matrices $\tilde{A}_i,\tilde{B}_i$, the $i$th worker node computes $\tilde{C}_i=\tilde{A}_i\tilde{B}_i$ and returns it to the master. Some workers may fail to respond or respond after the master recovers the product, we call such workers as stragglers.
For effective straggler mitigation, the master only receives the answers from some subset of workers to recover the product $AA^\top$
by using a class of decoding functions $\bm{d}=\{d_{\mathcal{R}}:\mathcal{R}\subseteq [N]\}$, i.e.,
$AA^\top=d_{\mathcal{R}}(\{\tilde{C}_i:i\in\mathcal{R}\})$ for some $\mathcal{R}$.

The recovery threshold of a computation strategy $(\bm{f,g,d})$, denoted by $R(\bm{f,g,d})$,
is defined as the minimum integer $k$ such that the master node can recover the product $AA^\top$ through any $k$ workers' answers for all possible data matrix $A$. Note that the recovery threshold represents the least number of answers that the master needs to collect for recovering the product $AA^\top$, so the recovery threshold is an important metric to measure the performance of a computation strategy and it is desired to be as small as possible.

In this paper, we aim to design a computation strategy with the minimum possible recovery threshold for the above mentioned computing problem. Since linear codes have low complexity in the encoding and decoding process with respect to the size of the input matrices among all possible computation strategies, we are interested in linear codes in the following.

\begin{definition}\label{def1} For the distributed matrix multiplication problem
of computing $AA^\top$ using $N$ workers, we say that a computation
strategy is a linear code with parameters $m,p$, if there
is a partitioning of the input matrices $A\in \mathbb{F}^{\mu\times\nu}$ and $A^\top\in \mathbb{F}^{\nu\times\mu}$, where each
matrix is evenly divided into the following sub-matrices of equal sizes
\begin{align}
&A=\begin{pmatrix}A_{0,0}&A_{0,1}&\cdots&A_{0,p-1}\\
A_{1,0}&A_{1,1}&\cdots&A_{1,p-1}\\
\vdots&\vdots&\ddots&\vdots\\
A_{m-1,0}&A_{m-1,1}&\cdots&A_{m-1,p-1}
\end{pmatrix},
&A^\top=\begin{pmatrix}
A^\top_{0,0}&A^\top_{1,0}&\cdots&A^\top_{m-1,0}\\
A^\top_{0,1}&A^\top_{1,1}&\cdots&A^\top_{m-1,1}\\
\vdots&\vdots&\ddots&\vdots\\
A^\top_{0,p-1}&A^\top_{1,p-1}&\cdots&A^\top_{m-1,p-1}
\end{pmatrix},\;\label{eqB}
\end{align}
such that the encoding functions $f_i,g_i$, $1 \le i \le N$, have the following form:
$$f_i(A)=\sum_{s=0}^{m-1}\sum^{p-1}_{k=0}a_{i,s,k}A_{s,k},~~ g_i(A^\top)=\sum_{s=0}^{m-1}\sum^{p-1}_{k=0}b_{i,s,k}A^\top_{s,k},
 $$
 for some tensors $a,b\in \mathbb{F}^{N\times m\times p}$, and the
decoding function for each recovery subset $\mathcal{R}$ can be written as the following form:
$$ C_{j,k}=\sum_{l=0}^{p-1}A_{j,l}A^\top_{k,l}=\sum_{i\in\mathcal{R}}\tilde{C}_ic_{i,j,k}
 $$
for some tensor
$c\in\mathbb{F}^{|\mathcal{R}|\times m\times m}$, where $A_{j,l}, A_{k,l} \in\mathbb{F}^{\frac{\mu}{m}\times \frac{\nu}{p}}$,
$0\leq j,k\leq m-1$ and $0 \le l \le p-1$.
 \end{definition}
\begin{definition} For the distributed matrix multiplication problem of  computing $AA^\top$ using $N$ worker nodes with matrix partitioning parameters $m,p$, the optimal linear recovery threshold, denoted by $R^*_{\rm linear}$,
is defined  as the minimum achievable recovery threshold among all linear codes. That is, $ R^*_{\rm linear}\triangleq \min_{({\bm f,\bm g,\bm d})\in\mathcal{L}}R({\bm f,\bm g,\bm d}),$ where $\mathcal{L}$ is the set of all linear codes for general matrix partitioning with parameters $m$ and $p$.
 \end{definition}

The goal of this paper is to design computation strategies with a low recovery threshold and to characterize the linear optimal recovery threshold $R^*_{\rm linear}$.

\subsection{Folded Polynomials}

In this subsection, we introduce the definition of folded polynomials and give a method to reconstruct such polynomials by using Alon's combinatorial nullstellensatz theorem \cite{Alon99}.
It's worth noting that the definition of folded polynomials presented here are due to the special structure of $AA^{\top}$-type matrix multiplication.
Actually, the decoding of our code relies on reconstructing such folded polynomials, and this is why our code is called folded polynomial code.

\begin{definition}
For a given polynomial $f(x)\in \mathbb{F}[x]$ with degree $n$, and a given set $\mathcal{G}=\{g_i(x): 1\leq i\leq k\}\subseteq \mathbb{F}[x]$  of $k$ linearly independent polynomials of degree $\leq n$, we say $f(x)$ is a $k$-terms folded polynomial with respect to $\mathcal{G}$, if there exist $a_i\in \mathbb{F}, i\in[1:k]$ such that $f(x)=\sum^k_{i=1}a_ig_i(x)$.  Moreover,  we call $g_i(x)\in \mathcal{G}$ a term of $f(x)$ and $k$  is called the number of terms of $f(x)$ with respect to $\mathcal{G}$.
\end{definition}

Next we recall the  Alon's combinatorial nullstellensatz theorem \cite{Alon99}.
\begin{theorem}\label{lem}(Combinatorial Nullstellensatz \cite{Alon99}) Let $\mathbb{F}$ be a field and let $f(x_1,x_2,...,x_n)\in\mathbb{F}[x_1,x_2,...,x_n]$ be a polynomial with ${\rm deg}(f)=\sum_{i=1}^nt_i$,
and the coefficient of $\prod_{i=1}^nx_{i}^{t_i}$ in $f$ is non-zero,  where  $t_i \ge 0$. Let $\mathcal{S}_i\subseteq \mathbb{F}$, for $i\in [1:n]$, be subsets with $|\mathcal{S}_i|>t_i$ and $\mathcal{S}:= \mathcal{S}_1 \times \mathcal{S}_2\times \cdots \times \mathcal{S}_n\in \mathbb{F}^n$. Then, there exists
$(r_1,r_2,...,r_n)\in \mathcal{S}$ such that $f(r_1,r_2,...,r_n)\neq 0$.
\end{theorem}

Based on the above combinatorial nullstellensatz theorem, we can build the following lemma, which gives a method to reconstruct folded polynomials.
\begin{lemma}\label{lem0}
For $j=1,2$, let  $\mathcal{G}_j=\{g_{i,j}(x)$, $i\in[1:k_j]\}$ be a set of $k_j$ linearly independent polynomials with degree $\leq n$. Suppose   
 $f_j(x)=\sum^{k_j}_{i=1}a_{i,j}g_{i,j}(x)$ with degree $n$ is a $k_j$-terms folded polynomial with respect to $\mathcal{G}_j$, $j\in[1:2]$.
Let $N>k_1>k_2$. If $|\mathbb{F}|>\sum^2_{j=1}\binom{N-1}{k_j-1}n$, then there exist $N$ points $\{\alpha_i: i\in[1:N]\}$ in $\mathbb{F}$ such that $f_j(x)$ can be reconstructed by using  any $k_j$ evaluations  $\{f_j(\alpha_{i_s}): s\in[1:k_j]\}$. That is, for $j=1,2$, the coefficients $a_{i,j}$'s can be uniquely determined by $\{f_j(\alpha_{i_s}): s\in[1:k_j]\}$.
\end{lemma}
\begin{proof}
The proof is given in Appendix \ref{apd00}.
\end{proof}

\begin{remark} The decoding process of folded polynomial codes relies on reconstructing several folded polynomials, which is heavily dependent on the selection of evaluation points in field $\mathbb{F}$ (see Section IV). Here Lemma \ref{lem0} provides a criterion for choices of evaluation points.
\end{remark}

\section{Main Results}

In this section, we present the main results of our paper in the following theorems.

\begin{theorem}\label{thm3}
For the distributed matrix multiplication problem of computing $AA^\top$ using $N$ worker nodes for general matrix partitioning with parameters $m$ and $p$, there exists a linear computation strategy,
referred to as folded  polynomial codes(FPC), achieving the recovery threshold
$$
R_{\rm FP}=\binom{m+1}{2}+\frac{(p-1)(2m^2-m+1)+\chi(m)\chi(p)}{2}
$$
if $|\mathbb{F}|>\left(\binom{N-1}{R_{\rm FP}-1}+\binom{N-1}{R_{\rm FP}-1-m}\right)(m^2p+p-2),{\rm char} \mathbb{F}\neq2$, or  $|\mathbb{F}|>\binom{N-1}{R_{\rm FP}-1}(m^2p+p-2),{\rm char} \mathbb{F}=2$, where $\chi(x)=\frac{1+(-1)^x}{2}$ for $x \in \mathbb{N}$. In particular, for $m=1$ such folded polynomial codes can be explicitly constructed as long as $|\mathbb{F}|>2N$.
\end{theorem}

\begin{remark} The proof of Theorem \ref{thm3} will be given in Section IV after introducing FP codes.
 The key to reducing the recovery threshold is that folded polynomials can be reconstructed by using less successful worker servers when computing $AA^\top$. It is noted that
FP codes degenerate into EP codes when they perform the task of $AB$ matrix multiplication.
\end{remark}

Next we show the second result, which characterizes a lower bound of the recovery threshold for general matrix partitioning with parameters $m$ and $p$. In particular, folded polynomial code is optimal for the case $m=1$  in all linear codes  over $\mathbb{R}$.
\begin{theorem}\label{thm2}
For the distributed matrix multiplication problem of computing $AA^\top$  over $\mathbb{R}$ using $N$ worker nodes with general matrix partitioning parameters $m$ and $p$, we have
\begin{equation}\label{eqmr1}
R^*_{\rm linear}\geq \min\{N,mp\}.
\end{equation}
Moreover, if $m=1$, then $R^*_{\rm linear}=R_{\rm FP}=p$.
  \end{theorem}
\begin{proof} Note that when $m=1$, $R^*_{\rm linear}=R_{\rm FP}=p$ can be directly obtained by Theorem \ref{thm3} and (\ref{eqmr1}), so it is sufficient to prove  (\ref{eqmr1}), that is, if $R^*_{\rm linear}<N$, then $R^*_{\rm linear}\geq mp$.

Suppose $\mathcal{L}^*$ is a linear computation strategy with the recovery threshold $R^*_{\rm linear}$  over $\mathbb{R}$. By Definition \ref{def1}, there exist tensors $a,b \in \mathbb{R}^{N\times m\times p}$ and the decoding functions ${\bm d}\triangleq\{d_{\mathcal{K}}:\mathcal{K}\subseteq[1:N],|\mathcal{K}|=R^*_{\rm linear}\}$ such that
\begin{align}\label{eqh4}
d_{\mathcal{K}}\left(\big\{(\sum_{s',k'}a_{i,s',k'}A_{s',k'})(\sum_{s,k}b_{i,s,k}A^\top_{s,k}
)\big\}_{i\in{\mathcal{K}}}\right)=AA^\top
\end{align}
for any input matrix $A\in\mathbb{R}^{\mu\times\nu}$ and any subset $\mathcal{K}$ of $R^*_{\rm linear}$ worker nodes.

For simplicity, we first introduce a vectorization operator ${\rm Vec}$, which maps a matrix $M\in\mathbb{R}^{m\times n}$ to a row vector in $ \mathbb{R}^{mn}$ whose entries are successively drawn from the matrix row by row.
Then choose $A_{i,k}=\lambda_{i,k}A_c$ for all  $\Lambda=(\lambda_{i,k})\in\mathbb{R}^{m\times p}$ in (\ref{eqh4}), where $ A_c\in\mathbb{R}^{\frac{\mu}{m}\times\frac{\nu}{p}}$ is  some nonzero matrix.
For $i\in[1:N]$, let ${\small {\bm a}_i={\rm Vec}((a_{i,j,k})_{j\in[1:m],k\in[1:p]})}$, ${\small{\bm b}_i={\rm Vec}((b_{i,j,k})_{j\in[1:m],k\in[1:p]})}$ and  ${\small{\bm \lambda}={\rm Vec}(\Lambda)}$.
Using these notations, we rewrite (\ref{eqh4}) as
$d_{\mathcal{K}}\big(\{({\bm a}_i\cdot{\bm \lambda}^\top)({\bm b}_i\cdot{\bm \lambda}^\top
)A_cA^\top_c\}_{i\in{\mathcal{K}}}\big)=AA^\top.$
Now we prove that the rank of $\{{\bm a}_i:i\in\mathcal{K}\}$ is $mp$. Otherwise, there exists a non-zero matrix ${\Lambda}_0\in \mathbb{R}^{p\times m}$ such  that for all $i\in\mathcal{K}$, ${\bm a}_i\cdot {\rm Vec}(\Lambda_0)^\top=0$.  Thus, for all $\lambda\in \mathbb{R}$ and $A=\Lambda_0\otimes A_c$,
\begin{align*}
AA^\top&=d_{\mathcal{K}}\big(\{({\bm a}_i\cdot{\rm Vec}(\Lambda_0)^\top)({\bm b}_i\cdot{\rm Vec}(\Lambda_0)^\top
)A_cA^\top_c\}_{i\in{\mathcal{K}}}\big)\\
&\stackrel{(a)}{=}d_{\mathcal{K}}\big(\{({\bm a}_i\cdot{\lambda\rm Vec}(\Lambda_0)^\top)({\bm b}_i\cdot{\lambda{\rm Vec}(\Lambda_0)^\top}
)A_cA^\top_c\}_{i\in{\mathcal{K}}}\big)\\
&=\lambda^2AA^\top,
\end{align*}where $(a)$ follows from  ${\bm a}_i\cdot {\rm Vec}(\Lambda_0)^\top=\lambda{\bm a}_i\cdot {\rm Vec}(\Lambda_0)^\top=0$.  Note that $A=\Lambda_0\otimes A_c\neq 0$, that is, ${\rm rank}(A)>0$. Moreover, ${\rm rank}(AA^\top)={\rm rank}(A)$ over $\mathbb{R}$, then $AA^\top\neq 0$, which implies that $\lambda^2=1$, a contradiction. Thus, $R^*_{\rm linear }=|\mathcal{K}|\geq mp$. \end{proof}

\section{Folded Polynomial Codes}

In this section, we prove Theorem \ref{thm3} by giving the folded polynomial codes. We start with two illustrating examples, and then present the folded polynomial codes with general parameters $m, p$ and use Lemma \ref{lem0} to ensure the decodability of such codes. At last, the complexity analysis of folded polynomial codes is given.

\subsection{Illustrating Examples}
\begin{example}\label{ex1}FP codes for $m=1,p=2$ with $R_{\rm FP}=2$. \\
Formally, we give precise encoding and decoding processes about how this FP code works. Let $|\mathbb{F}|\geq 2N$ and $\alpha_1,\alpha_2,...,\alpha_N$ are $N$ distinct elements in $\mathbb{F}$ with $\alpha_i\alpha_j\neq 1$ for $1\leq i,j\leq N$. The data matrices are divided into
 {\small$A=\begin{pmatrix}A_0&A_1\end{pmatrix},~~A^\top=\begin{pmatrix}A^\top_0\\A^\top_1\end{pmatrix},
 $} and then $C=AA^\top=\sum_{i=0}^1A_iA^\top_i$. 
 \begin{itemize}
\item \textbf {Encoding:} 
 Define the encoding functions $f_A(x)=A_0+A_1 x$ and $g_A(x)=A^\top_0 x+A^\top_1.$
 The master node computes $f_A(\alpha_i), g_A(\alpha_i)$ and sends them to the $i$th worker node for all $i\in[1:N]$.
\item\textbf {Worker node:} 
 After receiving the matrices $f_A(\alpha_i), g_A(\alpha_i)$, the $i$th worker node computes the matrix product $\tilde{C}_{i}=f_A(\alpha_i)g_A(\alpha_i)$ and returns it to the master node.
\item\textbf {Decoding:}
Now we show the master node can decode and get the desired product $AA^\top$ when receiving answers from any $2$ worker nodes.
We explain this as follows.
 \end{itemize}

We observe that for each $\alpha\in\{\alpha_1,\alpha_2,\cdots,\alpha_N\}$,
  $f_A(\alpha)g_A(\alpha)=B_1+AA^\top\alpha+B_1^\top\alpha^2,$
where $B_1=A_0A^\top_1$. Consider the $(i,j)$-entry and $(j,i)$-entry of $f_A(\alpha)g_A(\alpha)$, then we have
   \begin{equation}\label{eqmat2}
   \begin{split}
  &[f_A(\alpha)g_A(\alpha)]_{i,j}=[B_1]_{i,j}+[AA^\top]_{i,j}\alpha+[B_1]_{j,i}\alpha^2,\\
  &[f_A(\alpha)g_A(\alpha)]_{j,i}=[B_1]_{j,i}+[AA^\top]_{i,j}\alpha+[B_1]_{i,j}\alpha^2.
  \end{split}
   \end{equation}
According to (\ref{eqmat2}),  it has
  $[f_A(\alpha)g_A(\alpha)]_{i,i}=[B_1]_{i,i}(1+\alpha^2)+[AA^\top]_{i,i}\alpha$.
For any $2$-subset $\{\alpha_{s_1},\alpha_{s_2}\}$,
 \begin{equation}\label{eqmat3}\begin{split}
   &\det\begin{pmatrix}
   \alpha_{s_1}& 1+\alpha^2_{s_1}\\
   \alpha_{s_2}&1+\alpha^2_{s_2}\
   \end{pmatrix}
   =(\alpha_{s_2}-\alpha_{s_1})(\alpha_{s_2}\alpha_{s_1}-1)
   \neq 0,\end{split}
   \end{equation}
which implies that the master node can use any $2$ answers $\{\tilde{C}_{s_1},\tilde{C}_{s_2}\}$ to obtain the diagonal elements $[AA^\top]_{i,i}$. As to the off-diagonal elements $\{[AA^\top]_{i,j}:i\neq j\}$, according to (\ref{eqmat2}), the master node knows 
  $[f_A(\alpha_1)g_A(\alpha_1)]_{i,j}-[f_A(\alpha_1)g_A(\alpha_1)]_{j,i}=([B_1]_{i,j}-[B_1]_{j,i})(1-\alpha_1^2)$, and then $[B_1]_{i,j}-[B_1]_{j,i}$ can be recovered by $\alpha^2_1\neq1$. Define $[h(x)]_{i,j}=-([B_1]_{i,j}-[B_1]_{j,i})x^2$, the master node can compute the evaluations of $[f_A(x)g_A(x)-h(x)]_{i,j}=[B_1]_{i,j}(1+x^2)+[AA^\top]_{i,j}x$
at points  $\alpha_{s_1}, \alpha_{s_2}$. By (\ref{eqmat3}), one can directly recover the off-diagonal elements $\{[AA^\top]_{i,j}:i\neq j\}$. Hence $R_{\rm FP}=2$ for the case $m=1,p=2$.
  \end{example}
  \begin{remark} Note that the recovery threshold of MatDot codes with the same matrix partitioning in example 1 is $3$, which is larger than that of our FPC. This implies that there exists a more efficient computation strategy in the case $m=1,p=2$.
  \end{remark}
\begin{remark}\label{rmk1} In fact, when ${\rm char}\mathbb{F}\neq 2$, according to (\ref{eqmat3}), one can easily recover $[AA^\top]_{i,j}$ for  $i\neq j$ by using two evaluations of  the folded polynomial $$[f_A(x)g_A(x)]_{i,j}+[f_A(x)g_A(x)]_{j,i}=([B_1]_{i,j}+[B_1]_{j,i})(1+x^2)+2[AA^\top]_{i,j}x$$  with respect to $\{1+x^2,x\}$.
\end{remark}

In the following, we present another example for $m=2$.
\begin{example}\label{ex2}FP codes for $m=2,p=2,{\rm char}\mathbb{F}=2$ with ${R}_{\rm FP}=7$. \\
In this case, suppose $|\mathbb{F}|>8\binom{N-1}{6}$. Let $\alpha_1,\alpha_2,\cdots,\alpha_N$ be $N$ distinct undetermined evaluation points in $\mathbb{F}$, which are specified later. The data matrices are evenly divided as follows:
\begin{equation*}
 A=\begin{pmatrix}A_{0,0}&A_{0,1}\\A_{1,0}&A_{1,1}\end{pmatrix},~~A^\top=\begin{pmatrix}A^\top_{0,0}&A^\top_{1,0}\\A^\top_{0,1}&A^\top_{1,1}\end{pmatrix}.
 \end{equation*}
Then $C=AA^{\top}=(C_{i,j})_{i,j=0,1}$. The master node needs to compute $C_{i,j}=\sum_{k=0}^2A_{i,k}A^\top_{j,k}$ for all $i,j\in[0:1]$. Note
$C_{1,0}=C_{0,1}^{\top}$ by the symmetry of $C$.

\begin{itemize}
\item \textbf {Encoding:} 
Let $f_A(x)=\sum^1_{i=0}\sum^{1}_{j=0}A_{i,j}x^{2i+j}$ and $g_A(x)=\sum^1_{i=0}\sum^{1}_{j=0}A^\top_{i,j}x^{4i+1-j}$. Then the  master node sends $f_A(\alpha_i)$ and $g_A(\alpha_i)$  to the $i$th worker node.

\item \textbf {Worker node:}  The $i$th worker node computes $\tilde{C}_i=f_A(\alpha_i)g_A(\alpha_i)$ and returns the result to the master node.

\item \textbf{ Decoding:}
Now we show that the master node can recover the desired matrix $AA^\top$ by using answers from any $7$ worker nodes. We explain this as follows.
\end{itemize}

At first, by a precise computation, one can get
   \begin{eqnarray*}
   f_A(x)g_A(x)&=&C_{0,0}x+C_{1,1}x^7+(A_{1,1}A_{0,0}^\top+A_{0,0}A_{1,1}^\top)x^4+C_{1,0}x^3+C^\top_{1,0}x^5+\\
   && A_{0,0}A^\top_{0,1}+A_{0,1}A^\top_{0,0}x^2+A_{1,0}A_{0,1}^\top x^2+A_{0,1}A_{1,0}^\top x^6+A_{1,0}A_{1,1}^\top x^6+A_{1,1}A_{1,0}^\top x^8.
   \end{eqnarray*} 
Then consider the $(i,j)$-entry of $f_A(x)g_A(x)$, it has
   \begin{eqnarray}\label{eqmex2}
  [f_A(x)g_A(x)]_{i,j}
  &=&[C_{0,0}]_{i,j}x+[C_{1,1}]_{i,j}x^7+([A_{1,1}A_{0,0}^\top+A_{0,0}A_{1,1}^\top]_{i,j})x^4+\\\nonumber
 & &[C_{1,0}]_{i,j}x^3+[C_{1,0}]_{j,i}x^5+[A_{0,0}A^\top_{0,1}]_{i,j}+
   [A_{0,0}A^\top_{0,1}]_{j,i}x^2+\\\nonumber
  & &[A_{1,0}A_{0,1}^\top]_{i,j}x^2+[A_{1,0}A_{0,1}^\top]_{j,i}x^6+[A_{1,0}A_{1,1}^\top]_{i,j}x^6+[A_{1,0}A_{1,1}^\top]_{j,i}x^8. \nonumber
   \end{eqnarray}
Moreover, the $i$th diagonal entry of $f_A(x)g_A(x)$ is
   \begin{eqnarray}\label{eqmex3}
  [f_A(x)g_A(x)]_{i,i}
  &=&[C_{0,0}]_{i,i}x+[C_{1,1}]_{i,i}x^7+[C_{1,0}]_{i,i}(x^3+x^5)+\\ \nonumber
  &&[A_{0,0}A^\top_{0,1}]_{i,i}(1+x^2)+[A_{1,0}A_{0,1}^\top]_{i,i}(x^2+x^6)+[A_{1,0}A_{1,1}^\top]_{i,i}(x^6+x^8), \nonumber
   \end{eqnarray}
   where the term of $x^4$ is eliminated because $A_{0,0}A_{1,1}^\top=(A_{1,1}A_{0,0}^\top)^{\top}$ and $[A_{1,1}A_{0,0}^\top+A_{0,0}A_{1,1}^\top]_{i,i}=0$ since ${\rm char}\mathbb{F}=2$.
   Since $C_{0,0}$, $C_{1,1}$, $A_{1,1}A_{0,0}^\top+A_{0,0}A_{1,1}^\top$ are symmetric, then sum up the $(i,j)$-entry and $(j,i)$-entry of $f_A(x)g_A(x)$, it has
   \begin{eqnarray*}
   [f_A(x)g_A(x)]_{i,j}+[f_A(x)g_A(x)]_{j,i}
  &=&([C_{1,0}]_{i,j}+[C_{1,0}]_{j,i})(x^3+x^5)\\&&+([A_{0,0}A^\top_{0,1}]_{i,j}+[A_{0,0}A^\top_{0,1}]_{j,i})(1+x^2)\\
   &&+([A_{1,0}A_{0,1}^\top]_{i,j}+[A_{1,0}A_{0,1}^\top]_{j,i})(x^2+x^6)\\
   &&+([A_{1,0}A_{1,1}^\top]_{i,j}+[A_{1,0}A_{1,1}^\top]_{j,i})(x^6+x^8).
   \end{eqnarray*}
   Define $\Omega=\{x,x^7,x^3+x^5,1+x^2,x^2+x^6,x^6+x^8\}$. It is easy to verify that all polynomials of $\Omega$ are linearly independent over $\mathbb{F}$, so $[f_A(x)g_A(x)]_{i,i}$  is  a $6$-terms folded polynomial with respect to $\Omega$ and  $[f_A(x)g_A(x)]_{i,j}+[f_A(x)g_A(x)]_{j,i}$ is a $4$-terms folded polynomial with respect to $\Omega\setminus\{x,x^7\}$, respectively.

To recover the diagonal elements $\{[C_{s,t}]_{i,i}\!:\!0\leq s\!\leq\!\! t\!\leq\!\! 1\}$, the master needs to reconstruct the $6$-terms folded polynomial $[f_A(x)g_A(x)]_{i,i}$.
As to the off-diagonal elements $\{[C_{s,t}]_{i,j}:0\leq s,t\leq 1\}$ with $i\neq j$,
suppose the master node  can  recover the folded polynomial $[f_A(x)g_A(x)]_{i,j}+[f_A(x)g_A(x)]_{j,i}$, then as in example \ref{ex1}, it uses all coefficients of $[f_A(x)g_A(x)]_{i,j}+[f_A(x)g_A(x)]_{j,i}$ to construct an auxiliary polynomial $[h(x)]_{i,j}$. That is, let \begin{eqnarray}\label{eqaux}
[h(x)]_{i,j}&=&([C_{1,0}]_{i,j}+[C_{1,0}]_{j,i})x^5+([A_{0,0}A^\top_{0,1}]_{i,j}+[A_{0,0}A^\top_{0,1}]_{j,i})x^2+\\
&&
([A_{1,0}A_{0,1}^\top]_{i,j}+[A_{1,0}A_{0,1}^\top]_{j,i})x^6+ ([A_{1,0}A_{1,1}^\top]_{i,j}+[A_{1,0}A_{1,1}^\top]_{j,i})x^8. \nonumber
\end{eqnarray}
Then, we use $[h(x)]_{i,j}$ to build a folded polynomial whose coefficients contain the non-diagonal entries $\{[C_{s,t}]_{i,j}:0\leq s,t\leq 1\}$. That is, 
\begin{eqnarray*}
[f_A(x)g_A(x)]_{i,j}+[h(x)]_{i,j}&=&[C_{0,0}]_{i,j}x+[C_{1,1}]_{i,j}x^7+[C_{1,0}]_{i,j}(x^3+x^5)\\
&&+([A_{1,1}A_{0,0}^\top]_{i,j}+[A_{1,1}A_{0,0}^\top]_{j,i})x^4+
[A_{0,0}A^\top_{0,1}]_{i,j}(1+x^2)\\
&&+[A_{1,0}A_{0,1}^\top]_{i,j}(x^2+x^6)+[A_{1,0}A_{1,1}^\top]_{i,j}(x^6+x^8)
\end{eqnarray*} 
is a $7$-terms folded polynomial with respect to $\Omega\cup\{x^4\}$, which is because  polynomials in $\Omega\cup\{x^4\}$ are linearly independent over $\mathbb{F}$.
  To obtain $\{[C_{s,t}]_{i,j}:0\leq s,t\leq 1\}$, the master node needs to reconstruct two folded polynomials $[f_A(x)g_A(x)]_{i,j}+[f_A(x)g_A(x)]_{j,i}$ and $[f_A(x)g_A(x)]_{i,j}+[h(x)]_{i,j}$ by using $7$ evaluations of them. 
  
  Hence, for decoding the matrix $AA^\top$, the master node needs to recover three folded polynomials $[f_A(x)g_A(x)]_{i,i},[f_A(x)g_A(x)]_{i,j}+[f_A(x)g_A(x)]_{j,i}$ and $[f_A(x)g_A(x)]_{i,j}+[h(x)]_{i,j}$ by using $7$ evaluations of them. Moreover, one can find that every term of $[f_A(x)g_A(x)]_{i,i}$ and $[f_A(x)g_A(x)]_{i,j}+[f_A(x)g_A(x)]_{j,i}$ is also a term of $[f_A(x)g_A(x)]_{i,j}+[h(x)]_{i,j}$,  which implies that if $[f_A(x)g_A(x)]_{i,j}+[h(x)]_{i,j}$ can be recovered by using answers from any $7$ worker nodes, so do $[f_A(x)g_A(x)]_{i,i}$ and $[f_A(x)g_A(x)]_{i,j}+[f_A(x)g_A(x)]_{j,i}$. 
  A sufficient condition of decoding the matrix  $AA^\top$ is that there are $N$ points
$\{\alpha_j:j\in[1:N]\}$such that the folded polynomial $[f_A(x)g_A(x)]_{i,j}+[h(x)]_{i,j}$ can be reconstructed by any $7$ evaluations of it.  By Lemma \ref{lem0} and $|\mathbb{F}|>8\binom{N-1}{6}$,  the above condition holds.  Thus, $R_{\rm FP}=7$.
\end{example}

\begin{remark} In Example \ref{ex2}, our FP codes also work when ${\rm char}\mathbb{F}\neq 2$. That is, the diagonal elements of $\{[C_{s,t}]_{i,i}:1\leq s, t\leq 2\}$ can be recovered by using the $7$-terms folded polynomial $[f_A(x)g_A(x)]_{i,i}$  and  the off-diagonal elements of $\{[C_{s,t}]_{i,j}:1\leq s\leq t\leq 2\}$ can be recovered by using the $7$-terms folded polynomial $[f_A(x)g_A(x)]_{i,j}+[f_A(x)g_A(x)]_{j,i}$ and the $4$-terms folded polynomial $[f_A(x)g_A(x)]_{i,j}-[f_A(x)g_A(x)]_{j,i}$.
\end{remark}

Through observing the above examples, one can find that the key to decoding the FP codes is to recover some folded polynomials, say $[f_A(x)g_A(x)]_{i,i},[f_A(x)g_A(x)]_{i,j}\pm[f_A(x)g_A(x)]_{j,i}$ and $[f_A(x)g_A(x)]_{i,j}+[h(x)]_{i,j}$, where $[h(x)]_{i,j}$ is an auxiliary polynomial if necessary. The recovery threshold is determined by the number of terms of these folded polynomials. Next, we construct the FP codes for general matrix partitioning, and give an explicit construction for the case $m=1$.

\subsection{Folded Polynomial Codes for general $m\geq 1$}\label{subsec2}

Now we describe a general  folded  polynomial code that achieves a recovery threshold $R_{\rm FP}=\binom{m+1}{2}+\frac{(p-1)(2m^2-m+1)+\chi(m)\chi(p)}{2}$ for all possible $m\geq 1,p\geq 1$. First, the matrices $A$ and $A^\top$ are evenly divided into $m \times p$ and $p\times m$ sub-matrices according to  (\ref{eqB}). Then the master needs to compute $AA^{\top}=(C_{k,s})_{k,s\in[0:m-1]}$ where $C_{k,s}=\sum^{p-1}_{i=0}A_{k,i}A^\top_{s,i}$ for $k,s\in[0:m-1]$.
Suppose
$f_A(x)=\sum_{k=0}^{m-1}\sum^{p-1}_{j=0}A_{k,j}x^{kp+j},$    $g_A(x)=\sum_{s=0}^{m-1}\sum^{p-1}_{i=0}A^\top_{s,i}x^{smp+p-1-i},$ and $\alpha_1,\cdots,\alpha_N$ are $N$ distinct evaluation points in $\mathbb{F}$, which are specified later.
Then, {\small\begin{equation}\begin{split}\label{eqAB}
 f_A(x)g_A(x)
&=\sum_{k=0}^{m-1}\sum_{s=0}^{m-1}\sum^{p-1}_{j=0}\sum^{p-1}_{i=0}A_{k,j}A^\top_{s,i}x^{smp+kp+j+p-1-i}\\
&=\sum_{k=0}^{m-1}\sum_{s=0}^{m-1}\underbrace{C_{k,s}x^{smp+kp+p-1}}_{i=j}
+\sum_{k=0}^{m-1}\sum_{s=0}^{m-1}\sum_{t=1}^{p-1}(
\underbrace{B_{k,s,t}x^{smp+kp+p-1+t}}_{ i<j,\ t=j-i}
+\underbrace{B^\top_{k,s,t}x^{kmp+sp+p-1-t})}_{i>j,\ t=i-j},
\end{split}\end{equation}}where $C^\top_{k,s}=C_{s,k}$  and $B_{k,s,t}=\sum_{i=0}^{p-1-t}A_{k,i+t}A^\top_{s,i}$ for all $0\leq k,s\leq m-1$ and $1\leq t\leq p-1$. FP codes  are described as follows.
 \begin{itemize}
 \item \textbf {Encoding:} The master node computes $f_A(\alpha_i), g_A(\alpha_i)$ and sends them to the $i$th worker for all $i\in[1:N]$.
 \item\textbf {Worker node:} After receiving the matrices $f_A(\alpha_i), g_A(\alpha_i)$, the $i$th worker node computes the matrix product $f_A(\alpha_i)g_A(\alpha_i)$ and returns it to the master node.
\item\textbf {Decoding:} Upon receiving the  answers from any $\binom{m+1}{2}+\frac{(p-1)(2m^2-m+1)+\chi(m)\chi(p)}{2}$ worker nodes, the master must decode the desired product $AA^\top$.
 \end{itemize}

 The feasibility of decoding process will be shown in the proof of Theorem \ref{thm3} in the next subsection. 
 
 \subsection{The Proof of Theorem \ref{thm3}}
 
 Before presenting the proof of Theorem \ref{thm3}, we first recall the definition of folded  polynomials by a toy example. Let $f(x)=(x^4+1)+3(x+x^3)+2x^2\in \mathbb{F}[x]$.
Since $\{x^4+1,x+x^3,x^2\}$ are linearly independent over $\mathbb{F}$, 
 $f(x)$ is a $3$-terms folded polynomial with respect to them. 
In general,  for a polynomial $f(x)$ with the form $f(x)=\sum^k_{i=1}a_ig_i(x)$, where $g_i(x),i\in[1:k]$ are linearly independent over $\mathbb{F}$, we just call it a $k$-terms folded polynomial, if the polynomial set $\{g_i(x): 1\leq i\leq k\}$ is clear from the context.

In the following, we give some intuition and ideas about how to perform the decoding process of FP codes and then present the details of the proof of Theorem \ref{thm3}.

 To recover $C_{k,s}$ in (\ref{eqAB}), we use the symmetry of $AA^\top$ to construct some folded polynomials, i.e., $[f_A(x)g_A(x)]_{i,i},[f_A(x)g_A(x)]_{i,j}\pm[f_A(x)g_A(x)]_{j,i}$ and $[f_A(x)g_A(x)]_{i,j}+[h(x)]_{i,j}$, just as in Example \ref{ex1} and Example \ref{ex2}. 
 Specifically, considering the $(i,j)$-entry and $(j,i)$-entry of $f_A(x)g_A(x)$,  one can  obtain{\small\begin{eqnarray}\label{eqplus}&&[f_A(x)g_A(x)]_{i,j}+[f_A(x)g_A(x)]_{j,i}  \nonumber\\&=&
    \sum^{m-1}_{k=0}2[C_{k,k}]_{i,j}x^{kmp+kp+p-1} +\sum_{0\leq s< k\leq m-1}([C_{k,s}]_{i,j}+[C_{k,s}]_{j,i})(x^{smp+kp+p-1}+x^{kmp+sp+p-1})  \nonumber\\  \nonumber
    &&+\sum_{0\leq s,k\leq m-1}\sum_{t=1}^{p-1}([B_{k,s,t}]_{i,j}+[B_{k,s,t}]_{j,i})(x^{smp+kp+p-1+t}+x^{kmp+sp+p-1-t}) \\ 
    &=&\sum^{m-1}_{k=0}[C_{k,k}]_{i,j}g^{(1)}_{k,k,0}(x)+\sum_{0\leq s< k\leq m-1}([C_{k,s}]_{i,j}+[C_{k,s}]_{j,i})g^{(1)}_{k,s,0}(x)\\ 
    &&+\sum_{0\leq s,k\leq m-1}\sum_{t=1}^{p-1}([B_{k,s,t}]_{i,j}+[B_{k,s,t}]_{j,i})g^{(1)}_{k,s,t}(x) \nonumber
    \end{eqnarray}}
   and
{\small\begin{eqnarray}\label{eqminus}
    &&[f_A(x)g_A(x)]_{i,j}-[f_A(x)g_A(x)]_{j,i} \nonumber\\ \nonumber
    &=&\sum_{0\leq s< k\leq m-1}([C_{k,s}]_{i,j}-[C_{k,s}]_{j,i})(x^{smp+kp+p-1}-x^{kmp+sp+p-1})\\ \nonumber
    &&+\sum_{0\leq s,k\leq m-1}\sum_{t=1}^{p-1}([B_{k,s,t}]_{i,j}-[B_{k,s,t}]_{j,i})(x^{smp+kp+p-1+t}-x^{kmp+sp+p-1-t}) \\ 
    &=&\sum_{0\leq s< k\leq m-1}([C_{k,s}]_{i,j}-[C_{k,s}]_{j,i})g^{(2)}_{k,s,0}(x)+\sum_{0\leq s,k\leq m-1}\sum_{t=1}^{p-1}([B_{k,s,t}]_{i,j}-[B_{k,s,t}]_{j,i})g^{(2)}_{k,s,t}(x), 
    \end{eqnarray}}where $g^{(j)}_{k,s,t}(x)=x^{smp+kp+p-1+t}+(-1)^{j-1}x^{kmp+sp+p-1-t}$ for  $0\leq s,k\leq m-1,0\leq t\leq p-1, j =1, 2$.
For simplicity, define $\Omega_1\triangleq
  \{g^{(1)}_{k,s,t}(x):(k,s,t)\in[0:m-1]\times[0:m-1]\times[1:p-1]\}$ and  $\Omega_2\triangleq\{g^{(2)}_{k,s,t}(x):(k,s,t)\in[0:m-1]\times[0:m-1]\times[1:p-1]\}$. 
Then $[f_A(x)g_A(x)]_{i,j}+[f_A(x)g_A(x)]_{j,i}$ in (\ref{eqplus}) is a folded polynomial with respect to $\{g^{(1)}_{k,k,0}(x): k\in[0:m-1]\}\cup\{g^{(1)}_{k,s,0}(x): 0\leq s<k\leq m-1\}\cup \Gamma_1$, and  $[f_A(x)g_A(x)]_{i,j}-[f_A(x)g_A(x)]_{j,i}$ in (\ref{eqminus}) is a folded polynomial with respect to $\{g^{(2)}_{k,s,0}(x): 0\leq s<k\leq m-1\}\cup \Gamma_2$, where $\Gamma_i$ is a basis of ${\rm Span}(\Omega_i)$ for $i=1, 2$. 
    
Actually, once recovered the folded polynomials $[f_A(x)g_A(x)]_{i,j}+[f_A(x)g_A(x)]_{j,i}$ in (\ref{eqplus}) and $[f_A(x)g_A(x)]_{i,j}-[f_A(x)g_A(x)]_{j,i}$ in (\ref{eqminus}), we can obtain $[C_{k,k}]_{i,j}$ and $[C_{k,s}]_{i,j}$ when ${\rm char}\mathbb{F}\neq 2$. 
 As to ${\rm char}\mathbb{F}=2$,  (\ref{eqplus}) and (\ref{eqminus}) are the same equation. In this case, we recover the desired  matrix $AA^\top$ with the help of an auxiliary polynomial $[h(x)]_{i,j}$ as in (\ref{eqaux}) of Example 2. That is, we use  $[h(x)]_{i,j}$ to construct a folded polynomial $[f_A(x)g_A(x)]_{i,j}+[h(x)]_{i,j}$, whose coefficients contain $[C_{k,s}]_{i,j}$. Moreover, $[C_{k,s}]_{i,i}$ can be recovered by considering the folded polynomial $[f_A(x)g_A(x)]_{i,i}$.

 Hence, the key to  proving Theorem 6 is to determine the number of terms of the above folded polynomials by Lemma \ref{lem0}, which
 strictly depend on the dimensions of  ${\rm Span}(\Omega_1)$ and ${\rm Span}(\Omega_2)$.
  To this end,  we need some analysis on the polynomials in $\Omega_1$ and $\Omega_2$.

Observe that, in $\Omega_1$, the polynomials are of the form $x^{smp+kp+p-1+t}+x^{kmp+sp+p-1-t}=g^{(1)}_{k,s,t}(x)$ for all $(k,s,t)\in[0:m-1]\times[0:m-1]\times[1:p-1]$. One can find that any two polynomials of $\Omega_1$ share at most one monomial, and  each monomial of $g^{(1)}_{k,s,t}(x)\in \Omega_1$  appears in at most two polynomials of $\Omega_1$. Thus, once there are some polynomials in $\Omega_1$ that are linearly dependent, they must form a loop in terms of these monomials. 
  Actually, $\Omega_1$ can be partitioned into many disjoint chains, and the adjacent polynomials in each chain share exactly one monomial. These chains may end with a loop or not. If it is not a loop, we call it a single chain for simplicity, the polynomials in the single chain are linearly independent. The polynomials in $\Omega_2$ can be analysed similarly.

By the above observation, we characterize these chains as follows. For simplicity, we first define a map $\phi$ from $[0:m-1]\times[0:m-1]\times[1:p-1]\setminus \{(m-1,m-1,t):t\in[1:p-1]\}$ to  $[0:m-1]\times[0:m-1]\times[1:p-1]$: 
$$\phi(k,s,t)=
\begin{cases}
(s+1,0,p-t), ~\ \mathrm{if} \ k=m-1;\\
(s,k+1,p-t), ~\ \mathrm{if} \  0\leq k\leq m-2. 
\end{cases}
$$ 
One can easily verify that the first term of $g^{(i)}_{k,s,t}(x)$ is equal to the second term of  $g^{(i)}_{\phi(k,s,t)}(x)$ for all $(k,s,t)\in [0:m-1]\times[0:m-1]\times[1:p-1]\setminus \{(m-1,m-1,t):t\in[1:p-1]\}$ and $i\in[1:2]$. Thus $\{\phi^j(k,s,t): j\in\mathbb{N}\}$ gives a polynomial chain starting from $g^{(i)}_{k,s,t}(x)$, where these $\phi^j(k,s,t)$'s are considered as ordered. Next we use this map to construct all loops and single chains of $\Omega_1$ and $\Omega_2$.

Through a precise calculation and analysis, we find that all loops are exactly those chains starting from $(k,s,t)$ with $k=m-1$.
For $(s,t)\in [0:m-2]\times[1:p-1]$, define
$\mathcal{Y}_{s,t}\triangleq\{\phi^j(m-1,s,t):j\in \mathbb{N}\}$, which is a loop. However, different pairs $(s,t)$ may give the same loop.
After a simple calculation, one can find that there are at most two pairs relating one loop. Specifically, one can verify $(m-1,s,t)=\phi^{2(m-1)}(m-1,s,t)=\phi^{2s+1}(m-1,m-2-s,p-t)$, which implies that $\mathcal{Y}_{s,t}= \mathcal{Y}_{m-2-s,p-t}$ for $(s,t)\in [0:m-2]\times[1:p-1]$ and 
 $(m-1,\frac{m}{2}-1,\frac{p}{2})=\phi^{m-1}(m-1,\frac{m}{2}-1,\frac{p}{2})$ if $m,p$ are even. Moreover, 
$\mathcal{Y}_{s,t}=\{\phi^j(m-1,s,t):j\in[0:2(m-1)-1]\}$.
 Note that $\mathcal{Y}_{\frac{m}{2}-1,\frac{p}{2}}=\{\phi^j(m-1,s,t):j\in[0:m-2]\}$ appears only once if exists. Removing the repetitions of loops, we can obtain the set of all pairs $(s,t)$ inducing different loops $\mathcal{Y}_{s,t}$:
  $$\mathcal{D} =\begin{cases}
\{(s,t):0\leq s\leq \frac{m-3}{2},t\in[1:p-1]\} &\text{~ if~}  m \text{~is~odd},\\
 \{(s,t):0\leq s\leq \frac{m-4}{2},t\in[1:p-1]\}\cup\{(\frac{m-2}{2},t):t\in[1:\frac{p-1+\chi(p)}{2}]\}  &\text{~Otherwise}.
 \end{cases}$$
 
 Next we present all the single chains of $\Omega_1$ and $\Omega_2$.  Define $\mathcal{Y}_{0,0,t}\triangleq\{\phi^j(0,0,t):j\in \mathbb{N}\}$ for $t\in[1:p-1]$. Since $\phi^{2(m-1)}(0,0,t)=(m-1,m-1,t)$ and $\phi(m-1,m-1,t)$ is not defined, then $\mathcal{Y}_{0,0,t}$ is a single chain. Note the single chains $\mathcal{Y}_{0,0,t}, t\in[1:p-1]$ are all disjoint. Let  
 $\mathcal{Y}_{0,0}=\cup_{t\in[1:p-1]}\mathcal{Y}_{0,0,t}$. 
 One can verify that $\bigsqcup_{(s,t)\in\mathcal{D}\cup\{(0,0)\}}\mathcal{Y}_{s,t}=[0:m-1]\times[0:m-1]\times [1:p-1]$, which implies $\mathcal{Y}_{0,0,t}, t\in[1:p-1]$ are all the single chains.

 In the following, we determine the dimensions of ${\rm Span}(\Omega_1)$ and ${\rm Span}(\Omega_2)$ by calculating the dimension of subspace spanned by polynomials in each loop and single chain. This is illustrated in Lemma \ref{lem10}.

 \begin{lemma}\label{lem10} Using above notations,  then \begin{itemize}
\item[(i)]For $(a,b)\in\mathcal{D}\setminus\{(\frac{m}{2}-1,\frac{p}{2})\}$,  $\dim {\rm Span}(\{g^{(j)}_{k,s,t}(x):(k,s,t)\in\mathcal{Y}_{a,b}\})=|\mathcal{Y}_{a,b}|-1, j\in[1:2]$, and
       \begin{align}
       g^{(1)}_{m-1,a,b}(x)
       =&\sum^{2m-3}_{j=1}(-1)^{j-1}g^{(1)}_{\phi^j(m-1,a,b)}(x),
    \,\label{eq29}\\
       g^{(2)}_{m-1,a,b}(x)
       =&-\sum^{2m-3}_{j=1}g^{(2)}_{\phi^j(m-1,a,b)}(x). \label{eq29'}
       \end{align}
   \item[(ii)] If $(\frac{m}{2}-1,\frac{p}{2})\in\mathcal{D}$,  then
   $\dim{\rm Span}(\{g^{(1)}_{k,s,t}(x):(k,s,t)\in\mathcal{Y}_{\frac{m}{2}-1,\frac{p}{2}}\})=|\mathcal{Y}_{\frac{m}{2}-1,\frac{p}{2}}|-\delta_2({\rm char}\mathbb{F})$, and $\dim{\rm Span}\{g^{(2)}_{k,s,t}(x):(k,s,t)\in\mathcal{Y}_{\frac{m}{2}-1,\frac{p}{2}}\}=m-2$.  Moreover, when ${\rm char}\mathbb{F}= 2$,
\begin{equation}\label{eq330}\begin{split}g^{(1)}_{m-1,\frac{m}{2}-1,\frac{p}{2}}=&\sum^{m-2}_{j=1}g^{(1)}_{\phi^j(m-1,\frac{m}{2}-1,\frac{p}{2})}(x)
       .\end{split}\end{equation}
\item[(iii)] For $j\in[1:2]$,
\begin{equation}\label{eq28}
   {\rm Span}({\Omega_j})=
  \bigoplus_{(a,b)\in\mathcal{D}\cup\{0,0\}}{\rm Span}\{g^{(j)}_{k,s,t}(x):(k,s,t)\in\mathcal{Y}_{a,b}\}.
  \end{equation}
  Moreover,
 $$\dim {\rm Span}(\Omega_{1})=\frac{(p-1)(2m^2-m+1)+\chi(m)\chi(p)(1-2\delta_2({\rm char}\mathbb{F}))}{2}$$
and $$\dim {\rm Span}(\Omega_{2}) =\frac{(p-1)(2m^2-m+1)-\chi(m)\chi(p)}{2},$$
where $\delta_2(\cdot)$ is a function from $\mathbb{N}$ onto $\{0,1\}$ and $\delta_2(x)=1$ if and only if $x=2$.
 \end{itemize}
 \end{lemma}
 \begin{proof}
 To maintain the fluency of this paper, the proof is given in Appendix \ref{apd3}.
 \end{proof}

\emph{\textbf{The proof of Theorem \ref{thm3}:}} The proof is divided into two cases.

 {\bf Case 1. ${\rm char}\mathbb{F}\neq 2,R_{\rm FP}=\binom{m+1}{2}+\frac{(p-1)(2m^2-m+1)+\chi(m)\chi(p)}{2}$}.\\
  By  (\ref{eqplus}), (\ref{eqminus}), and  (iii) in Lemma \ref{lem10},  $[f_A(x)g_A(x)]_{i,j}\pm[f_A(x)g_A(x)]_{j,i}$ can be rewritten as follows,
\begin{eqnarray}
    \label{eqFpm3'}
&&[f_A(x)g_A(x)]_{i,j}+[f_A(x)g_A(x)]_{j,i}\\&=&\sum_{k=0}^{m-1}[C_{k,k}]_{i,j}g^{(1)}_{k,k,0}(x) +\sum_{g(x)\in{\Gamma}_{1}}a_{g}g(x)
+ \sum_{0\leq s<k\leq m-1}([C_{k,s}]_{i,j}+[C_{k,s}]_{j,i})g^{(1)}_{k,s,0}(x), \nonumber
\end{eqnarray} 
\begin{eqnarray}
\label{eqFpm3}
[f_A(x)g_A(x)]_{i,j}-[f_A(x)g_A(x)]_{j,i}
=\sum_{0\leq s<k\leq m-1}([C_{k,s}]_{i,j}-[C_{k,s}]_{j,i})g^{(2)}_{k,s,0}(x)
+\sum_{g(x)\in{\Gamma}_{2}}b_{g}g(x),\end{eqnarray}where ${\Gamma}_{i}$ is a basis of  ${\rm Span}(\Omega_i)$ for $i\in[1:2]$, $a_g$ and $b_{g}$ are some elements in $\mathbb{F}$. Moreover,  one can verify that $\{g^{(j)}_{k,s,0}(x):0\leq k\leq s \leq m-1\}\cap{\rm Span}(\Omega_j)=\emptyset$ for $j\in\{1,2\}$, and $\{g^{(j)}_{k,s,0}(x):0\leq k\leq s \leq m-1\}$ are linearly independent over $\mathbb{F}$. Hence, the numbers of terms of {\small$[f_A(x)g_A(x)]_{i,j}+[f_A(x)g_A(x)]_{j,i}$}  and  {\small$[f_A(x)g_A(x)]_{i,j}-[f_A(x)g_A(x)]_{j,i}$} are {\small$\binom{m+1}{2}+\dim {\rm Span}(\Omega_1)$}  and  {\small$\binom{m}{2}+\dim {\rm Span}(\Omega_2)$}, respectively. 
Since ${\rm char}\mathbb{F}\neq 2$, $[C_{k,s}]_{i,j},[C_{k,s}]_{j,i}$ can be recovered from $[C_{k,s}]_{i,j}+[C_{k,s}]_{j,i}$ and $[C_{k,s}]_{i,j}-[C_{k,s}]_{j,i}$.
Similarly, the diagonal entries $[C_{k,s}]_{i,i}$ can be recovered by reconstructing the folded polynomial $[f_A(x)g_A(x)]_{i,i}$, which can be verified to have the same terms as $[f_A(x)g_A(x)]_{i,j}+[f_A(x)g_A(x)]_{j,i}$.
By Lemma \ref{lem0} and {\small$|\mathbb{F}|>(pm^2+p-2)\left(\binom{N-1}{\binom{m+1}{2}+\dim {\rm Span}(\Omega_1)-1}+\binom{N-1}{{\binom{m}{2}+\dim {\rm Span}(\Omega_2)-1}}\right),$} there always exist $N$ points in $\mathbb{F}$ such that the folded polynomials $[f_A(x)g_A(x)]_{i,j}\pm[f_A(x)g_A(x)]_{j,i}$ and $[f_A(x)g_A(x)]_{i,i}$ can be recovered from any $\binom{m+1}{2}+\dim {\rm Span}(\Omega_1)$ worker nodes' answers.
 Hence, $R_{\rm FP}=\binom{m+1}{2}+\dim{\rm Span}(\Omega_1)= \binom{m+1}{2}+\frac{(p-1)(2m^2-m+1)+\chi(m)\chi(p)}{2}$.

{\bf Case 2.} ${\rm char}\mathbb{F}=2,R_{\rm FP}=\binom{m+1}{2}+\frac{(p-1)(2m^2-m+1)+\chi(m)\chi(p)}{2}$. \\
 In this case, $\Omega_1=\Omega_2$. By (\ref{eq28}) in Lemma \ref{lem10}, polynomials in {\small$\bigcup_{(a,b)\in\mathcal{D}\cup{(0,0)}}\{g^{(1)}_{k,s,t}(x):(k,s,t)\in\mathcal{\tilde{Y}}_{a,b}\}$}  form a basis of ${\rm Span}(\Omega_1)$, where $\mathcal{\tilde{Y}}_{a,b}=\mathcal{Y}_{a,b}\setminus\{(m-1,a,b)\}$.
According to (\ref{eq29}) and (\ref{eq330}), 
{\small\begin{eqnarray*}
&&\sum_{0\leq k,s\leq m-1}\sum^{p-1}_{t=1}([B_{k,s,t}]_{i,j}+[B_{k,s,t}]_{j,i})g^{(1)}_{k,s,t}(x)\\&=&\sum_{(k,s,t)\in\mathcal{Y}_{0,0}}([B_{k,s,t}]_{i,j}+[B_{k,s,t}]_{j,i})g^{(1)}_{k,s,t}(x)+\\
&&\sum_{(a,b)\in\mathcal{D}}\sum_{(k,s,t)\in\mathcal{\tilde{Y}}_{a,b}}
([B_{k,s,t}]_{i,j}+[B_{m-1,a,b}]_{i,j}+[B_{k,s,t}]_{j,i}+[B_{m-1,a,b}]_{j,i})g^{(1)}_{k,s,t}(x).
\end{eqnarray*}}
Then (\ref{eqplus}) can be rewritten as
{\small\begin{equation}\label{eqFpm3''}
 \begin{split}   
&[f_A(x)g_A(x)]_{i,j}+[f_A(x)g_A(x)]_{j,i}\\
=&\sum_{0\leq s<k\leq m-1}([C_{k,s}]_{i,j}+[C_{k,s}]_{j,i})g^{(1)}_{k,s,0}(x)+ 
\sum_{(k,s,t)\in\mathcal{Y}_{0,0}}([B_{k,s,t}]_{i,j}+
[B_{k,s,t}]_{j,i})g^{(1)}_{k,s,t}(x) +\\
&\sum_{(a,b)\in\mathcal{D}}\sum_{(k,s,t)\in\mathcal{\tilde{Y}}_{a,b}}
([B_{k,s,t}]_{i,j}+[B_{m-1,a,b}]_{i,j}+[B_{k,s,t}]_{j,i}+[B_{m-1,a,b}]_{j,i})g^{(1)}_{k,s,t}(x), 
\end{split}
\end{equation}}hence $[f_A(x)g_A(x)]_{i,j}+[f_A(x)g_A(x)]_{j,i}$ is a $\binom{m}{2}+\frac{(p-1)(2m^2-m+1)+\chi(m)\chi(p)}{2}$-terms folded polynomial.
Once recovered this folded polynomial, the master node can
build an auxiliary polynomial
{\small\begin{equation*}\begin{split}
[h(x)]_{i,j}=&\sum_{0\leq s<k\leq m-1}([C_{k,s}]_{i,j}+[C_{k,s}]_{j,i})x^{kmp+sp+p-1}+
\sum_{(k,s,t)\in\mathcal{Y}_{0,0}}([B_{k,s,t}]_{i,j}+[B_{k,s,t}]_{j,i})x^{kmp+sp+p-1-t}\\
&+\sum_{(a,b)\in\mathcal{D}}\sum_{(k,s,t)\in\mathcal{\tilde{Y}}_{a,b}}
([B_{k,s,t}]_{i,j}+[B_{m-1,a,b}]_{i,j}+[B_{k,s,t}]_{j,i}+[B_{m-1,a,b}]_{j,i})x^{kmp+sp+p-1-t}.\end{split}
\end{equation*}}
According to (\ref{eqAB}), it has
{\small\begin{eqnarray}\label{eqFpm31}
[f_A(x)g_A(x)]_{i,j}+[h(x)]_{i,j}&=&\sum_{k=0}^{m-1}[C_{k,k}]_{i,j}x^{kmp+kp+p-1}+\sum_{0\leq s<k\leq m-1}[C_{k,s}]_{i,j}g^{(1)}_{k,s,0}(x)+\\ \nonumber
&&
\sum_{0\leq k,s\leq m-1}\sum^{p-1}_{t=1}[B_{k,s,t}]_{i,j}g^{(1)}_{k,s,t}(x)+
\\ \nonumber
&&\sum_{(a,b)\in\mathcal{D}}
([B_{m-1,a,b}]_{i,j}+[B_{m-1,a,b}]_{j,i})\sum_{(k,s,t)\in\mathcal{Y}_{a,b}}x^{kmp+sp+p-1-t}\\ \nonumber
 &\stackrel{(a)}{=}& \sum_{k=0}^{m-1}[C_{k,k}]_{i,j}x^{kmp+kp+p-1}+\sum_{0\leq s<k\leq m-1}[C_{k,s}]_{i,j}g^{(1)}_{k,s,0}(x)\\ \nonumber
 &&+\sum_{(a,b)\in\mathcal{D}\cup\{(0,0)\}}\sum_{(k,s,t)\in\mathcal{\tilde{Y}}_{a,b}}b_{k,s,t}g^{(1)}_{k,s,t}(x)\\ \nonumber
&&+\chi(m)\chi(p)([B_{m-1,\frac{m}{2}-1,\frac{p}{2}}]_{i,j}+[B_{m-1,\frac{m}{2}-1,\frac{p}{2}}]_{j,i})x^{(m-1)mp+\frac{m-1}{2}p-1},\nonumber
\end{eqnarray}}where $b_{k,s,t}$ are some coefficients in $\mathbb{F}$ for all $(k,s,t)\in\bigcup_{(a,b)\in\mathcal{D}}\mathcal{Y}_{a,b}$, and {\small$(a)$} comes from the fact that 
{\small\begin{align*}&\sum_{(k,s,t)\in\mathcal{Y}_{\frac{m}{2}-1,\frac{p}{2}}}x^{kmp+sp+p-1-t}
=x^{(\frac{m}{2}+1)p+\frac{p}{2}-1}\cdot\sum^{m-2}_{j=0}x^{jp(m+1)}
=\sum^{\frac{m}{2}-2}_{i=0}g^{(1)}_{\frac{m}{2}+i,i,\frac{p}{2}}(x)+x^{(m-1)mp+\frac{m-1}{2}p-1}, 
\end{align*}}and for $(a,b)\in \mathcal{D}\setminus\{(\frac{m}{2}-1,\frac{p}{2})\}$, {\small\begin{eqnarray*}
\sum_{(k,s,t)\in\mathcal{Y}_{a,b}}x^{kmp+sp+p-1-t}
&=&(x^{(m-a)p+b-1}+x^{p-1-b+(a+2)p})\cdot\sum^{m-2}_{j=0}x^{jp(m+1)}\\
&=&\sum^{m-2-a}_{i=0}g^{(1)}_{a+1+i,i,p-b}(x)+\sum^{a-1}_{i=0}g^{(1)}_{i,m-a+i,p-b}(x)\in{\rm Span}(\Omega_1),\end{eqnarray*}} here {\small\begin{align*}\mathcal{Y}_{a,b}=&\{(m-1,a,b),(a+1,0,p-b)\}\cup\{(m-1-a+i,i,b),(i,m-a+i,p-b):i\in[0:a-1]\}\\&\cup
 \{(i,a+2+i,b),(a+2+i,i+1,p-b):i\in[0: m-3-a]\}.\end{align*}}
Hence, $[f_A(x)g_A(x)]_{i,j}+[h(x)]_{i,j}$ is a $\binom{m+1}{2}+\frac{(p-1)(2m^2-m+1)+\chi(m)\chi(p)}{2}$-terms folded polynomial.

By (\ref{eqFpm3''}) and (\ref{eqFpm31}), one can find that every term of $[f_A(x)g_A(x)]_{i,j}+[f_A(x)g_A(x)]_{j,i}$ is also a term of $[f_A(x)g_A(x)]_{i,j}+[h(x)]_{i,j}$, which implies that if $[f_A(x)g_A(x)]_{i,j}+[h(x)]_{i,j}$ can be recovered by some $t$ evaluation  points, so does $[f_A(x)g_A(x)]_{i ,j}+[f_A(x)g_A(x)]_{j,i}$. Then, a sufficient condition of decoding $AA^\top$  for the case ${\rm char}\mathbb{F}=2$ is that
there exist $N$ points $\alpha_1,\cdots,\alpha_N\in\mathbb{F}$ such that the master node can reconstruct the folded polynomials $[f_A(x)g_A(x)]_{i,j}+[h(x)]_{i,j}$ from any $\binom{m+1}{2}+\frac{(p-1)(2m^2-m+1)+\chi(m)\chi(p)}{2}$ worker nodes' answers. Actually, this condition is guaranteed by Lemma \ref{lem0} and $|\mathbb{F}|>(pm^2+p-2)\binom{N-1}{\binom{m+1}{2}+\frac{(p-1)(2m^2-m+1)+\chi(m)\chi(p)}{2}}$. Hence, $R_{\rm FP}=\binom{m+1}{2}+\frac{(p-1)(2m^2-m+1)+\chi(m)\chi(p)}{2}.$ $\hfill\blacksquare$

According to the decoding process in Theorem \ref{thm3}, the field size required in the folded polynomial code is $\mathcal{O}\left((pm^2+p-2)\big(\binom{N-1}{R_{\rm FP}-1}+\binom{N-1}{R_{\rm FP}-m-1}\big)\right)=\mathcal{O}\left((pm^2+p-2)N^{ R_{\rm FP}}\right)$. To reduce the  field size, we present an explicit folded polynomial code  with $|\mathbb{F}|>2N$ for the case of $m=1$ by Corollary \ref{coro1} in the next subsection.

 \subsection{Folded Polynomial Codes for $m=1$}
 In this subsection, we first give two lemmas and then use them to present an explicit folded polynomial code for the case $m=1$, which achieves a recovery threshold $R_{\rm FP}=p$ with $|\mathbb{F}|>2N$.

\begin{lemma}\label{lem1} Suppose $\beta_1,\beta_2,...,\beta_n$ are $n$ distinct elements in $
\mathbb{F}$ and $\beta_i\beta_j\neq 1$ for all $1\leq i, j\leq n$. Assume $g_1(x)=x^{n-1}$ and $g_i(x)=x^{n-i}+x^{n-2+i}$ for $i\in[2:n]$. Then,
$\det {\rm M}(\Omega,\Gamma)
=\prod_{1\leq i<j\leq n}(\beta_j-\beta_i)(\beta_i\beta_j-1),$
where $\Omega=\{g_i(x):i\in[1:n]\}$ and $\Gamma=\{\beta_j:j\in[1:n]\}$.
\end{lemma}
\begin{proof}
The proof is shown in Appendix \ref{apd2}.
\end{proof}
\begin{lemma}\label{lem2} Suppose $\beta_1,\beta_2,\cdots,\beta_n$ are $n$ distinct elements in $
\mathbb{F}$ and $\beta_i\beta_j\neq 1$ for all $1\leq i, j\leq n$.
Then,
$\det {\rm M}(\Omega,\Gamma)
=\prod_{1\leq i<j\leq n}(\beta_j-\beta_i)(\beta_i\beta_j-1),$
where $\Omega=\{x^{n+i}-x^{n-i}:i\in[1:n]\}$ and $\Gamma=\{\beta_j:j\in[1:n]\}$.
\end{lemma}
\begin{proof}
The proof is shown in Appendix \ref{apd1}.
\end{proof}
According to Lemma \ref{lem1} and Lemma \ref{lem2},  we are able to determine $N$ points $\alpha_1,\cdots,\alpha_N\in\mathbb{F}$ that are needed in the decoding process in \ref{subsec2}, and an explicit folded polynomial code for $m=1$  can be constructed as illustrated in the following corollary.
\begin{corollary}\label{coro1} Assume $|\mathbb{F}|>2N$,  $\alpha_1,\alpha_2,\cdots,\alpha_N$ are $N$ distinct elements in $\mathbb{F}$ and $\alpha_i\alpha_j\neq 1$, $1\leq i, j\leq N$. Then, a folded polynomial code can be  explicitly constructed over $\mathbb{F}$ for $m=1$, and its recovery threshold  achieves $p$.
\end{corollary}
\begin{proof}
In the case of $m=1$, (\ref{eqAB}) can be rewritten as
\begin{equation}\label{eqh}
f_A(x)g_A(x)=B_0x^{p-1}+\sum^{p-1}_{\ell=1}(B_{\ell}x^{p-1+\ell}+B^\top_{\ell}x^{p-1-\ell}),
\end{equation}
where $B_0=AA^\top$ and  $B_\ell=\sum^{p-1-\ell}_{k=0}A_{k+\ell}A^\top_{k}$, $1\leq \ell\leq p-1$.

In the following, we state that the master node can reconstruct $AA^\top$ from any $p$ worker nodes.
We first recover the diagonal elements $[AA^\top]_{i,i}=[B_0]_{i,i}$. According to (\ref{eqh}), 
$[f_A(x)g_A(x)]_{i,i}=[B_0]_{i,i}x^{p-1}+\sum^{p-1}_{\ell=1}[B_\ell]_{i,i}(x^{p-1+\ell}+x^{p-1-\ell}).$ 
By Lemma \ref{lem1},  ${\rm M}(\Omega_1,\Gamma)$ is invertible, where $\Omega_1=\{x^{p-1-\ell}+x^{p-1+\ell}:\ell\in[1:p-1]\}\cup\{x^{p-1}\}$ and $\Gamma$ is a $p$-subset of $\{\alpha_i:i\in[1:N]\}$. Hence,  $[AA^\top]_{i,i}$ can be easily recovered from any $p$ evaluations of $\{[f_A(\alpha_{i_s})g_A(\alpha_{i_s})]_{i,i}:s\in[1:p]\}$.

Next we decode the off-diagonal elements $[AA^\top]_{i,j}$ with $i<j$. It follows from (\ref{eqh})  that
$[f_A(x)g_A(x)]_{i,j}-[f_A(x)g_A(x)]_{j,i}=\sum^{p-1}_{\ell=1}([B_\ell]_{i,j}-[B_\ell]_{j,i})(x^{p-1+\ell}-x^{p-1-\ell}),$ which
 is a $(p-1)$-terms folded polynomial. Let $\Omega_2=\{x^{p-1+\ell}-x^{p-1-\ell}:\ell\in[1:p-1]\}$. By Lemma \ref{lem2},
 for any $p-1$-subset $\Gamma'$ of  $\Gamma$,
$\det{\rm M}(\Omega_2,\Gamma')\neq 0$, which implies that the master can recover the folded polynomial $[f_A(x)g_A(x)]_{i,j}-[f_A(x)g_A(x)]_{i,j}$, and further knows the coefficients $[B_\ell]_{i,j}-[B_\ell]_{j,i}$ for all $\ell\in[1:p-1]$. Define $[h(x)]_{i,j}=\sum^{p-1}_{\ell=1}([B_\ell]_{i,j}-[B_\ell]_{j,i}) x^{p-1-\ell}$, then the master node can compute evaluations of the following $p$-terms folded polynomial
$$[f_A(x)g_A(x)]_{i,j}+[h(x)]_{i,j}=[B_0]_{i,j}x^{p-1}+\sum^{p-1}_{\ell=1}[B_\ell]_{i,j}(x^{p-1+\ell} +x^{p-1-\ell})$$at points $\alpha_{i_1},\alpha_{i_2},...,\alpha_{i_p}$.
Hence, $[AA^\top]_{i,j}$ can be recovered  by the invertibility of ${\rm M}(\Omega_1,\Gamma)$. Thus, $R_{\rm FP}=p$ for $m=1$.  \end{proof}

\subsection{Complexity Analysis of Folded Polynomial Codes}
In this subsection, we present the computational complexity and communication cost for folded  polynomial codes.

{\bf Encoding:}  The master encodes sub-matrices of
sizes $\frac{\mu}{m}\times \frac{\nu}{p}$ and $\frac{\nu}{p}\times \frac{\mu}{m}$ and sends them to each worker node. Hence, the encoding process for matrices $A, A^\top$ can be viewed as evaluating two polynomials $f_{A}(x)$ and $g_{A}(x)$ of respective degrees $pm-1$ and $(m^2-m+1)p-1$ at $N$ points for $\frac{\mu\nu}{mp}$ times. By using the fast polynomial evaluation algorithms \cite{Modern13}, the  computational complexity of the encoding process is $\mathcal{O}\left(N\mu\nu\frac{\log^2(m^2 p)\log\log (m^2p)}{mp}\right)$.

{\bf Each worker's computational cost:} Each worker computes the product of
two matrices of size $\frac{\mu}{m}\times \frac{\nu}{p}$ and $\frac{\nu}{p}\times \frac{\mu}{m}$, which requires $\frac{\mu^2\nu}{m^2p}$ operations by using standard matrix multiplication algorithms. Hence, the computational complexity of each worker is $\mathcal{O}\left(\frac{\mu^2\nu}{m^2p}\right)$.

{\bf Decoding:} After receiving the fastest $R_{\rm FP}$ answers from workers $\Gamma=\{\alpha_{i_s}: s\in[1:R_{\rm FP}]\}$, the master reconstructs the desired matrix $AA^\top$. When ${\rm char}\mathbb{F}\neq 2$, the decoding process can be viewed as solving two linear systems whose equations have form $(\ref{eqFpm3})$ and $(\ref{eqFpm3'})$ respectively. Hence, the computational complexity  in the decoding process is {\small\begin{align*}
  \mathcal{O}\left(\binom{m}{2}\binom{\frac{\mu}{m}}{2}(R_{\rm FP}-m)+\binom{m+1}{2}\binom{\frac{\mu}{m}+1}{2}R_{\rm FP}+(R_{\rm FP}-m)^3+R^3_{\rm FP}\right)=\mathcal{O}\left(\frac{\mu(\mu+1)}{2}R_{\rm FP}+R^3_{\rm FP}\right).
\end{align*}}When ${\rm char}\mathbb{F}=2$, compared with the case of ${\rm char}\mathbb{F}\neq2$, the master has to additionally  compute the evaluations of $\{[h(x)]_{i,j}:1\leq i<j\leq \frac{\mu}{m}\}$ at $R_{\rm FP}$ points, which requires $\mathcal{O}\left(\binom{\frac{\mu}{m}}{2}R^2_{\rm FP}\right)$ operations. Thus, the computational complexity in decoding process is  $$\mathcal{O}\left(\frac{\mu(\mu+1)}{2}R_{\rm FP}+R^3_{\rm FP}+\delta_2({\rm char} \mathbb{F})\binom{\frac{\mu}{m}}{2}R^2_{\rm FP}\right).$$

{\bf Communication cost:} The communication cost consists of upload cost and download cost at the master node. In our scheme, the master node uploads a total of $2N\frac{\mu\nu}{mp}$ symbols to all $N$ worker nodes, and downloads $R_{\rm FP}\frac{\mu^2}{m^2}$ symbols from the fastest $R_{\rm FP}$ successful worker nodes. That is, the upload cost is $\mathcal{O}(N\frac{\mu\nu}{mp})$, and the download cost is  $\mathcal{O}(R_{\rm FP}\frac{\mu^2}{m^2})$. Thus, the total amount of communication cost is  $\mathcal{O}\left(N\frac{\mu\nu}{mp}+R_{\rm FP}\frac{\mu^2}{m^2}\right)$.

\section{Comparisons and Numerical Experiments}  
\subsection{Comparisons with previous works}
To the best of our knowledge, EP codes can be regarded as the state of art of previous works with respect to CDMM, because  it bridges the  extremes of Polynomial codes \cite{YMA17}, MatDot codes,  and PolyDot codes  \cite{MatDotIT20}. Moreover, generalized PolyDot (GPD) codes \cite{DCG18} and EP codes have similar  performance, but the former reduced the recovery threshold of PolyDot codes by a factor of $2$.  Since our strategy is designed to perform the task of computing $AA^\top$ for matrix partitioning with parameters $m,p$, we will compare our code with previous ones under the same task and the same parameters $m,p$. The comparisons are listed in Table \ref{tab0} and Table \ref{tab1}.
\begin{table}[ht]
\centering
\renewcommand\arraystretch{1.5}
\caption{Computation Strategies for the CDMM  problem of computing $AA^\top$ using $N$ worker nodes with parameters $p$ and $m=1$. }\label{tab0}
\begin{tabular}{|l|c|c|}
  \hline
 {\rm ~} & {\rm MatDot Codes}\cite{MatDotIT20}&{\rm FP  Codes}\\\hline
  {\rm Recovery~Thershold}&$2p-1$&$p$\\\hline
  {\rm Encoding~Complexity}&$\mathcal{O}\left(N\mu\nu\frac{\log^2 p\log\log p}{p}\right)$&$\mathcal{O}\left(N\mu\nu\frac{\log^2 p\log\log p}{p}\right)$\\\hline
   {\rm Decoding~Complexity}& $\mathcal{O}\left((2p-1)\frac{\mu(\mu+1)}{2}+(2p-1)^3\right)$&$\mathcal{O}\left(p\frac{\mu(\mu+1)}{2}+p^3+\delta_2({\rm char} \mathbb{F})\binom{\mu}{2}p^2\right)$\\\hline
  {\rm  Worker~Computation~Complexity} &$ \mathcal{O}\left(\frac{\mu^2\nu}{p}\right)$&$\mathcal{O}\left(\frac{\mu^2\nu}{p}\right)$\\\hline
{\rm Upload~Cost}&$\mathcal{O}\left(N\frac{\mu\nu}{p}\right)$&$\mathcal{O}\left(N\frac{\mu\nu}{p}\right)$\\\hline
{\rm Download~Cost}&$\mathcal{O}\left((2p-1)\mu^2\right)$&$\mathcal{O}\left(p\mu^2\right)$\\\hline
\end{tabular}
\end{table}
\begin{table}[ht]
\centering
\caption{Computation Strategies for the CDMM  problem of computing $AA^\top$ using $N$ worker nodes with arbitrary fixed parameters $p$ and $m$. }\label{tab1}
\renewcommand\arraystretch{1.5}
\begin{tabular}{|l|c|c|}
  \hline
 {\rm ~} & {\rm EP Codes}\cite{YMIT20} &{\rm FP Codes}\\\hline
  {\rm Recovery~Thershold ($R$)}&$pm^2+p-1$&$\binom{m+1}{2}+\frac{(p-1)(2m^2-m+1)+\chi(m)\chi(p)}{2}$\\\hline
  {\rm Encoding~Complexity}&$\mathcal{O}\left(N\mu\nu\frac{\log^2 (pm^2)\log\log (pm^2)}{mp}\right)$&$\mathcal{O}\left(N\mu\nu\frac{\log^2 (pm^2)\log\log (pm^2)}{mp}\right)$\\\hline
   {\rm Decoding~Complexity}& $\mathcal{O}\left(\frac{\mu(\mu+1)}{2}R+R^3\right)$&$\mathcal{O}\left(\frac{\mu(\mu+1)}{2}R+R^3+\delta_2({\rm char} \mathbb{F})\binom{\frac{\mu}{m}}{2}R^2\right)$\\\hline
  {\rm  Worker~Computation~Complexity} &$ \mathcal{O}\left(\frac{\mu^2\nu}{pm^2}\right)$&$\mathcal{O}\left(\frac{\mu^2\nu}{pm^2}\right)$\\\hline
{Upload Cost}&$\mathcal{O}\left(N\frac{\mu\nu}{mp}\right)$&$\mathcal{O}\left(N\frac{\mu\nu}{mp}\right)$\\\hline
{Download cost}&$\mathcal{O}\left(R\frac{\mu^2}{m^2}\right)$&$\mathcal{O}\left(R\frac{\mu^2}{m^2}\right)$\\\hline
\end{tabular}
\end{table}
For matrix partitioning with $m=1$ in Table \ref{tab0},  we have the following observations. Compared with the MatDot codes in\cite{MatDotIT20}, FP codes reduce the recovery threshold from $2p-1$ to $p$, 
reduce the download cost by a factor of $2$, and preserve the other performance unchanged. In particular, to ensure the decodability, we need the field size $|\mathbb{F}|\geq 2N$, which is almost the same as the requirement in MatDot codes.

 Table \ref{tab1} is for general parameters $m,p$. It can be seen that compared with EP codes, FP codes are better in terms of recovery threshold, download cost and decoding complexity for all parameters $m,p$ on condition that  the underlying  fields of the two code schemes are the same, while encoding complexity and woker computation complexity do not change. Particularly, when $p=1$, our FP code has the recovery threshold $\binom{m+1}{2}$, which reduces that of EP codes  by a factor of $2$. Furthermore, it should be pointed out that  the field size required in FP codes is $ \mathcal{O}\big((pm^2+p-2)N^{t}\big)$ for some $t$, which is larger than the field size $\mathcal{O}(N)$ in EP codes. This may lead to higher upload and download cost for FP codes when defining over a large field.  However, we claim that the field size condition given in this paper is just a sufficient condition to ensure the  decoding  process of FP codes. Actually, there is evidence that the field size can be further reduced. Therefore, building FP codes over a small field is an interesting problem  and we leave it as a future work.

When $\mathbb{F}=\mathbb{R}$, there are also many efficient approaches \cite{DA22,DAIT22},\cite{DAV21},\cite{
FCIT21,RT22} for CDMM. Compared with our approach, the methods, designed for  sparse matrices  multiplication and partial stragglers in \cite{DA22,DAIT22}, have a lower worker computation complexity. To keep numerical robustness, two methods are designed in \cite{FCIT21,RT22} with respect to different matrix partitioning, both of which give an upper bound on the condition number of the decoding matrices in the worst case.

\subsection{Numerical Experiments}
In this subsection, since FP codes for $m=1$ have an optimal linear recovery threshold over $\mathbb{R}$,  we do experiments using MatDot codes\cite{MatDotIT20}, OrthMatDot codes \cite{FCIT21} and our FP codes for matrix partitioning parameter $m=1$, and give comparisons on performance of these codes via exhaustive numerical experiments  in a computing node of MAGIC CUBE-III cluster (HPCPlus Platform in Shanghai Supercomputer Center, \cite{SSC}) that is equipped with  2 Intel Xeon Gold 6142 CPUs (2.6 GHz, 16 cores), 192 GB DDR4 memory, and 240 GB local SSD hard disk.
One can find the codes of all numerical experiments in \cite{softerfpc}.

{\bf Worker Computation Time \& Decoding Time}: In our simulation, the input matrix $A$ has size $12000\times 15000$, whose entries are chosen independently according to the standard Gaussian distribution  $\mathcal{N}(0,1)$.  The matrix $A$ is divided into $8$ column blocks, i.e., $m=1,p=8$. The output matrix $AA^\top$ is obtained by decoding all entries of it. In Table \ref{Tab3},
worker computation time is  measured by the average of computation time over all non-straggler worker nodes, 
and decoding time is the time spent by the master node  performing 
calculations during the decoding process. 
Moreover,  the  computation  complexity in worker nodes and master node are  determined by the number of  multiplication operations.

From Table \ref{Tab3}, we have the following observations. Worker computation time in our FP codes is almost the same as that of MatDot codes, but is longer than that of OrthMatDot codes, which is because that each worker node in OrthMatDot codes computes a $\tilde{A}_i\tilde{A}_i^\top$-type matrix multiplication.  As to decoding time at the master node, compared with OrthMatDot codes \cite{FCIT21}, our FP codes still reduce the decoding time despite the distributed computing system has more straggler nodes.
This coincides with the analysis on decoding
complexity  of the master  node, which is $\mathcal{O}\big(\mu^2R^2\big)=\mathcal{O}(3.24\times 10^{10})$ in OrthMatdot codes and $\mathcal{O}(1.15\times 10^{9})$ in FP codes.  Compared with  MatDot codes \cite{MatDotIT20}, the master node   in  FP codes  additionally  needs to perform some  addition operations (3.55 seconds), which is the reason  why the decoding  time  of FP codes is a litter longer than that of MatDot codes.  
\begin{table}
  \centering
  \caption{Comparison among three coded approaches in terms of
worker  computation time, decoding time for $N=18$
and the fixed matrix partition ($m=1,p=8$), where evaluation points are  selected as the $18$-dimensional Chebyshev points. 
}\label{Tab3}
\begin{tabular}{cccc}
\toprule
&{\rm MatDot Code\cite{MatDotIT20}}&{\rm OrthMatDot Code\cite{FCIT21}}&{\rm FP Code}\\
\midrule
{\rm No. of Stragglers} $s$&$3$&$3$&$10$\\
{\rm Worker Computation Complexity}&$\mathcal{O}\left(2.7\times 10^{11}\right)$&$\mathcal{O}\left(1.35\times 10^{11}\right)$&$\mathcal{O}\left(2.7\times 10^{11}\right)$\\
{\rm Worker Computation Time($sec.$)}&$7.78$&$4.57$&$7.77$\\
{\rm Decoding Complexity}&$\mathcal{O}\left(2.16\times 10^9\right)$&$\mathcal{O}\left(3.24\times10^{10}\right)$&$\mathcal{O}\left(1.15\times 10^9\right)$\\
{\rm Decoding Time$(sec.)$}&$1.80$&$12.86$&$4.55=1.10+3.55^*$\\
\bottomrule
\end{tabular}

 \vspace{0.2cm}
$^*$Decoding time in FP code consists of two parts: multiplication (1.10 sec.)  and addition (3.55 sec.).
\end{table}

{\bf Overall Computation Time}: We compare the above three approaches in terms of overall computation time to recover $AA^\top$ by carrying out some simulations for different number of straggler nodes, where the computing system has $18$ worker nodes. 
Note that overall computation time is the time required by the master node to receive the first $R$ answers, where $R$ is the recovery threshold. 
\begin{figure}[ht]
  \centering
  \includegraphics[width=6cm]{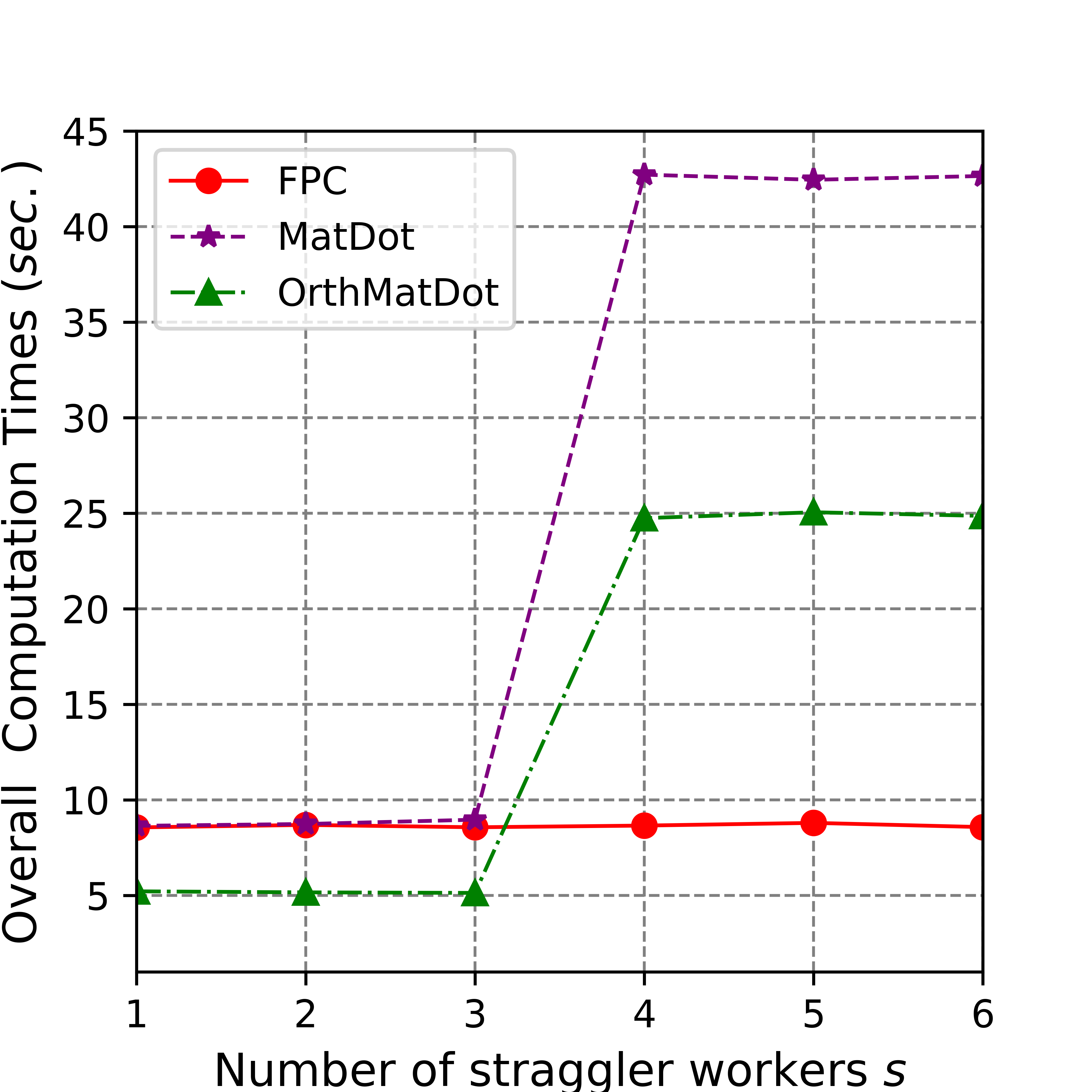}\\
  \caption{Comparison among three coded approaches in terms of overall computation time for different number of straggler nodes $s$. The computing system has $18$ worker nodes and the input matrix $A$ is evenly divided into $8$ column blocks and $A^\top$ is evenly divided into $8$ row blocks, so $R_{\rm FP}=8$, and the recovery threshold of both MatDot codes and OrthMatDot codes is $15$. The straggler workers are simulated in such a way so that they have one-fifth of the speed of the non-straggling workers.}\label{fig1}
\end{figure}
The results are presented in Fig.\ref{fig1}. One can find  that our proposed approach is significantly faster in terms of overall computation time in comparison to the other two coded approaches for different number of straggler workers. This is because that our method has a smaller recovery threshold for the same matrix partitioning and  can tolerate more straggler nodes as shown in Table \ref{tab0}.

\begin{figure}[ht]
  \centering
  \includegraphics[width=6cm]{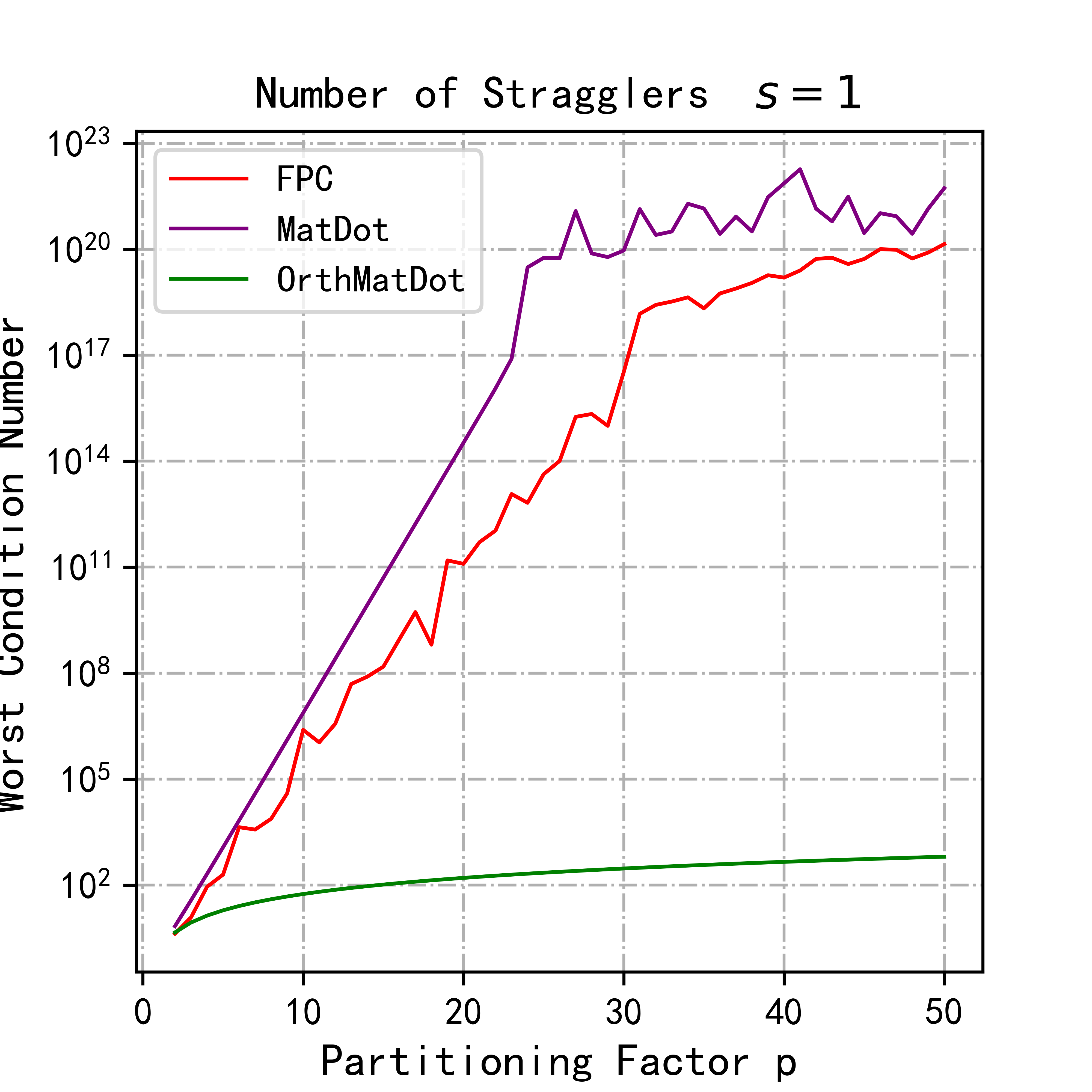}~\includegraphics[width=6cm]{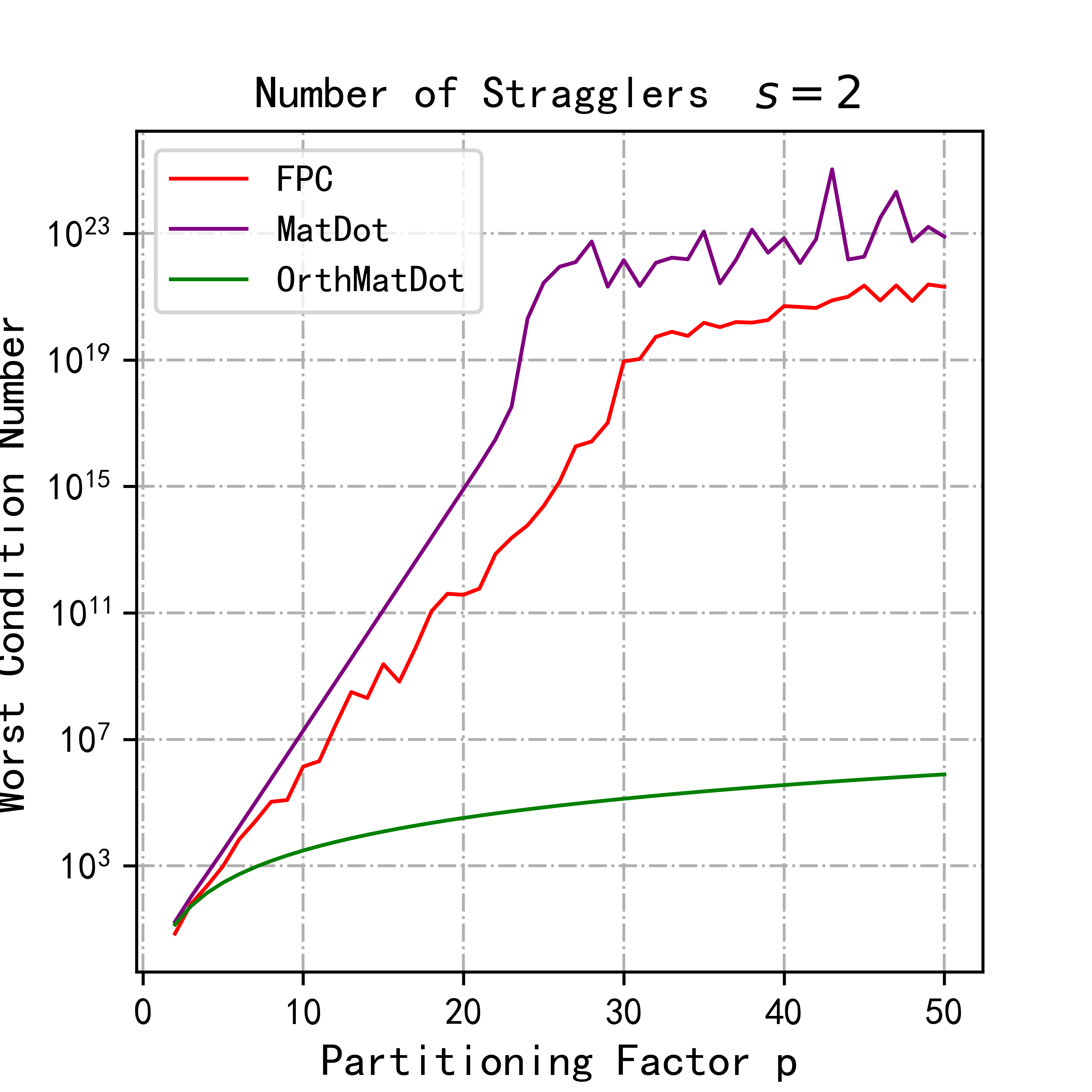}\\
  \caption{Comparison among three coded approaches in terms of
 worst condition number for $m=1$, different partitioning parameter $p$ and fixed number of straggler nodes $(s=1,2)$.}\label{fig2}
\end{figure}

{\bf Numerical Stability:} We know that the condition number of the associated decoding matrix is an important metric to measure the numerical stability of a linear computing strategy for CDMM, where the condition number of a matrix is defined by the ratio of maximum and minimum singular values.  Compared with OrthMatDot codes in Fig. \ref{fig2}, both our codes and MatDot codes have very high worst condition number, which implies they are numerically instable for fixed $s$ and different $p$. Here evaluation points in OrthMatDot codes and MatDot codes are the corresponding Chebyshev points, and evaluation points in our codes are given in the following manner: we randomly choose the evaluation points from the interval $(-1,1)$ and compute the worst condition number for 20 times, then choose those that have the minimum worst condition number. 
Due to the numerical instability of FP codes, it will be a future work to design an FP code which is numerically more stable.
\section{Conclusion}
In this paper, we investigate the distributed matrix multiplication problem of computing $AA^\top$ using $N$ worker nodes with parameters $m$ and $p$. We first design a linear computation strategy for general $m\geq 1,p\geq 1$, i.e., folded  polynomial code over $\mathbb{F}$,  which has a recovery threshold $R_{\rm FP}= \binom{m+1}{2}+\frac{(p-1)(2m^2-m+1)+\chi(m)\chi(p)}{2}$, where $\chi(x)=\frac{1+(-1)^x}{2}$ for $x \in \mathbb{N}$.  Then we derive a lower bound $R^*_{\rm linear}\geq \min\{N,mp\}$ on the recovery threshold for all linear computation strategies  over $\mathbb{R}$. In particular, our code for $m=1$ is explicitly constructed and has a recovery threshold $p$, achieving the lower bound. For general partitioning parameters $m\geq 2$,  our strategy is  constructed over a large  underlying field. Therefore, it is an interesting problem to construct folded polynomial codes for general $m\geq 2$ over a small field. Besides, the folded polynomial code over $\mathbb{R}$ is not numerically stable for
larger number of worker nodes. A future  work is  to design a numerically  stable scheme with the same recovery threshold as FP codes over $\mathbb{R}$.

\section{Acknowledgement}
The authors are grateful to the editor and the anonymous reviewers for their valuable comments which have highly improved the manuscript.

\appendices
\section{Proof of Lemma\ref{lem0}}\label{apd00}
To complete the proof of Lemma \ref{lem0}, we first prove the following claim.\\
{\textbf{Claim}}: If polynomials in $\mathcal{G}=\{g_i(x)\in\mathbb{F}[x]: \deg(g_i(x))\leq n,i\in[1:s]\}$ are linearly independent over $\mathbb{F}$ and $|\mathbb{F}|>n$, then there exist $s$ elements $\Gamma=\{\beta_j:j\in[1:s]\}$ in $\mathbb{F}$ such that $\det\mathcal{M}(\mathcal{G},\Gamma)\neq 0$.
\begin{proof}
This claim is proved by induction on $s$. For $s=1$, it is obviously true. Now let $s>1$ and suppose that the claim is true for $s-1$. By the induction hypothesis, there are $s-1$ elements $\Gamma_1=\{\beta_j:j\in[1:s-1]\}$ in $\mathbb{F}$ such that $\det\mathcal{M}(\mathcal{G}\setminus\{g_s(x)\},\Gamma_1)\neq 0$. Then we prove that there exist an element $\beta_s\in\mathbb{F}$ such that $\det\mathcal{M}(\mathcal{G},\Gamma_1\cup\{\beta_s\})\neq 0$. Otherwise, for every $\alpha\in\mathbb{F}$, $\det\mathcal{M}(\mathcal{G},\Gamma_1\cup\{\alpha\})=0$. Expand by the last column,  $\det\mathcal{M}(\mathcal{G},\Gamma_1\cup\{\alpha\})=\sum^s_{i=1}c_ig_i(\alpha)=0$ for all $\alpha\in\mathbb{F}$, where $c_i=(-1)^{s+i}\det\mathcal{M}(\mathcal{G}\setminus\{g_i(x)\},\Gamma_1)$ for $i\in[1:s]$. Combining with $|\mathbb{F}|>n$, $\sum^s_{i=1}c_ig_i(x)=0$. Note that $c_s=\det\mathcal{M}(\mathcal{G}\setminus\{g_s(x)\},\Gamma_1)\neq 0$, which implies that polynomials in $\mathcal{G}$ are linearly dependent over $\mathbb{F}$, a contradiction.
\end{proof}

\textbf{Proof of Lemma \ref{lem0}:}
In order to uniquely determine $a_{i,j}$'s, it is  sufficient to show that   the matrix
 ${\rm M}(\mathcal{G}_j,\mathcal{X}_j)$
  is invertable over $\mathbb{F}$, $j\in[1:2]$, for each $k_j$-subset $\mathcal{X}_j$  of $\{\alpha_i: i\in[1:N]\}$.

  Set $$G(x_1,x_2,\cdots,x_N)=\prod _{\substack{\Gamma_1\subseteq \Omega\\|\Gamma_1|=k_1}}\det {\rm M}(\mathcal{G}_1,\Gamma_1)\cdot \prod _{\substack{\Gamma_2\subseteq\Omega,\\ |\Gamma_2|=k_2}}\det {\rm M}(\mathcal{G}_2,\Gamma_2).$$
 Combined the fact that $g_{1,j}(x),\cdots$,$g_{k_j,j}(x)$ are linearly independent in $\mathbb{F}_{\leq n}[x]$ for each $j\in[1:2]$ and the 
 above \textbf {Claim}, we conclude that $\det {\rm M}(\mathcal{G}_j,\Gamma_j)$
is a nonzero polynomial for each $k_j$-subset $\Gamma_j$ of $\{x_i:i\in[1:N]\}$ and  the degree of each variable in $\det {\rm M}(\mathcal{G}_j,\Gamma_j)$ is at most $n$,  which implies that $G(x_1,x_2,\cdots,x_N)$ is also a nonzero polynomial in $\mathbb{F}[x_1,x_2,\cdots,x_N]$. Moreover,  the  variable $x_i$ actually appears in $\binom{N-1}{k_1-1}+\binom{N-1}{k_2-1}$ factors of $G(x_1,\cdots,x_N)$ for each $i\in[1:N]$,  hence the degree of each variable in $G(x_1,x_2,\cdots,x_N)$ is at most $\sum^2_{j=1}\binom{N-1}{k_j-1}n$. By
 Theorem \ref{lem} and $|\mathbb{F}|>\sum^2_{j=1}\binom{N-1}{k_j-1}n$, there are $N$ points $\{\alpha_i: i\in[1:N]\}$ in $\mathbb{F}$ such that $G(\alpha_1,\alpha_2,\cdots,\alpha_N)\neq 0$.     $\hfill\blacksquare$

\section{Proof of Lemma \ref{lem10}}
\label{apd3}

\begin{proof}Note that when $p=1$, $\Omega_1=\Omega_2=\emptyset$, the  statements in Lemma  \ref{lem10} clearly hold. W.L.O.G., we assume $p\geq 2$ in the following proof. 

For convenience,  we  first define two maps $\sigma_1,\sigma_2$ from $[0:m-1]\times[0:m-1]\times[0:p-1]$ to $\mathbb{N}$ by $\sigma_1(k,s,t)=smp+kp+p-1+t$ and $\sigma_2(k,s,t)=kmp+sp+p-1-t$, one can easily verify $\sigma_1,\sigma_2$ are injective.
Then $g^{(j)}_{k,s,t}(x)$ can be written as $g^{(j)}_{k,s,t}(x)=x^{\sigma_1(k,s,t)}+(-1)^{j-1}x^{\sigma_2(k,s,t)}$. According to the definition of $\phi$,  one can obtain $\sigma_1(k,s,t)=\sigma_2(\phi(k,s,t))$.

(i) Equations (\ref{eq29}) and (\ref{eq29'}) can be directly verified by a simple calculation. 
Next we show that for $j\in[1:2]$ and $(a,b)\in \mathcal{D}\setminus\{(\frac{m}{2}-1,\frac{p}{2})\}$, the polynomials $\{g^{(j)}_{k,s,t}(x):(k,s,t)\in\mathcal{Y}_{a,b}\setminus\{(m-1,a,b)\}\}$ are linearly independent over $\mathbb{F}$.
For $j\in[1:2]$, suppose $\sum_{i=1}^{2m-3}c_i g^{(j)}_{\phi^i(m-1,a,b)}(x)=0$ for some $c_{i}\in\mathbb{F}$. Then
{\small\begin{align*}
0&=\sum_{i=1}^{2m-3}c_i g^{(j)}_{\phi^i(m-1,a,b)}(x)\\
 &=\sum_{i=1}^{2m-3}c_i(x^{\sigma_1(\phi^i(m-1,a,b))}+(-1)^{j-1}x^{\sigma_2(\phi^i(m-1,a,b))}\\ 
 &=c_1(-1)^{j-1}x^{\sigma_1(m-1,a,b)}+\sum_{i=1}^{2m-4}(c_i+(-1)^{j-1}c_{i+1})x^{\sigma_1(\phi^i(m-1,a,b))} +c_{2m-3}x^{\sigma_1(\phi^{2m-3}(m-1,a,b))},
 \end{align*}}Since $\sigma_1$ is injective,  monomials in the above equation are different, which implies that these  coefficients are all zero, that is, $c_i=0$ for all $i\in[1:2m-3]$, $j\in\{1,2\}$. Hence, for $j\in[1:2]$, $\dim {\rm Span}(\{g^{(j)}_{k,s,t}(x):(k,s,t)\in\mathcal{Y}_{a,b}\})=2m-3.$

(ii) For $j\in[1:2]$, suppose  {\small$\sum_{i=0}^{m-2}c_{i} g^{(j)}_{\phi^i(m-1,\frac{m}{2}-1,\frac{p}{2})}(x)=0$} for some $c_{i}\in\mathbb{F}$. Similar to the proof of (i),  we rewrite the above equation as  a linear combination of some different monomials, i.e., 
\begin{align*}
 0&=\sum_{i=0}^{m-2}c_{i} g^{(j)}_{\phi^i(m-1,\frac{m}{2}-1,\frac{p}{2})}(x)\\
 &=\sum_{i=0}^{m-3}(c_i+(-1)^{j-1}c_{i+1})x^{\sigma_1(\phi^i(m-1,\frac{m}{2}-1,\frac{p}{2}))}+(c_{m-2}+(-1)^{j-1}c_0)x^{\sigma_1(\phi^{m-2}(m-1,\frac{m}{2}-1,\frac{p}{2}))}.
 \end{align*}Hence,
 $c_{i}=(-1)^jc_{i+1}$ for $i\in [0:m-3]$ and $c_{m-2}=(-1)^jc_{0}$. 
 That is,  $c_{i}=(-1)^{ij}c_{0}$ for all $i\in[1:m-2]$ and  $c_0=(-1)^{(m-1)j}=(-1)^jc_0$ since $m$ is even.  Then,  when $j=2$ or ${\rm char}\mathbb{F}=2$, one can find that $c_i=c_0$ for all $i\in[0:m-2]$. When $j=1$ and ${\rm char}\mathbb{F}\neq2$, it has $2c_0=0$, which implies that $c_{i}=0$ for all $i\in[0:m-2]$. Thus,  $\dim{\rm Span}\{g^{(1)}_{k,s,t}(x):(k,s,t)\in\mathcal{Y}_{\frac{m}{2}-1,\frac{p}{2}}\}=m-1-\delta_2({\rm char}\mathbb{F}),$ and $\dim{\rm Span}\{g^{(2)}_{k,s,t}(x):(k,s,t)\in\mathcal{Y}_{\frac{m}{2}-1,\frac{p}{2}}\}=m-2.$   
 Moreover, when ${\rm char}\mathbb{F}=2$, equation (\ref{eq330}) can be obtained by setting $c_0=1$.

(iii) For convenience, given a polynomial $g(x)=\sum^n_{i=0}a_ix^i$, denoting ${\rm Supp}(g(x))=\{x^i:a_i\neq 0,i\in[0:n]\}$ the support of $g(x)$. Since chains $\mathcal{Y}_{s,t}$ for $(s,t)\in\mathcal{D}\cup\{(0,0)\}$ are all disjoint, one can find that for $(s_1,t_1)\neq (s_2,t_2)\in \mathcal{D}\cup\{(0,0)\}$, $\{{\rm Supp}(g^{(j)}_{k,s,t}(x)):(k,s,t)\in\mathcal{Y}_{s_1,t_1}\}\cap \{{\rm Supp}(g^{(j)}_{k,s,t}(x)):(k,s,t)\in\mathcal{Y}_{s_2,t_2}\}=\emptyset.$
Hence, ${\rm Span}({\Omega_j})=
   \bigoplus_{(a,b)\in\mathcal{D}\cup\{0,0\}}{\rm Span}\{g^{(j)}_{k,s,t}(x):(k,s,t)\in\mathcal{Y}_{a,b}\}.$
 Similar as proof of (i), one can easily check that all polynomials $\{g^{(j)}_{k,s,t}(x):(k,s,t)\in\mathcal{Y}_{0,0}\}$ are linearly independent over $\mathbb{F}$, so $\dim {\rm Span}(\{g^{(j)}_{k,s,t}(x):(k,s,t)\in\mathcal{Y}_{0,0}\})=|\mathcal{Y}_{0,0}|=(2m-1)(p-1).$ Combining (i) and (ii), 
{\small\begin{eqnarray*}
\dim {\rm Span}(\Omega_1)
&=&|\mathcal{Y}_{0,0}|+\chi(m)\chi(p)(m-1-\delta_2({\rm char}\mathbb{F}))+\sum_{(s,t)\in\mathcal{D}\setminus\{\frac{m}{2}-1,\frac{p}{2}\}}(|\mathcal{Y}_{s,t}|-1)\\
&=&\frac{(p-1)(2m^2-m+1)+\chi(m)\chi(p)(1-2\delta_2({\rm char}\mathbb{F}))}{2},\\
\dim {\rm Span}(\Omega_2)
&=&|\mathcal{Y}_{0,0}|+\chi(m)\chi(p)(m-2)+\sum_{(s,t)\in\mathcal{D}\setminus\{\frac{m}{2}-1,\frac{p}{2}\}}|\mathcal{\tilde{Y}}_{s,t}| \\
&=&\frac{(p-1)(2m^2-m+1)-\chi(m)\chi(p)}{2}.
\end{eqnarray*}}
\end{proof}

\section{Proof of Lemma \ref{lem1}}
\label{apd2}

\begin{proof} Suppose $\Omega^{(j)}=\{x^{j-1}\}\cup\{x^{j-1+t}+x^{j-1-t}:t\in[1:j-1]\}$ and $\Gamma^{(j)}=\{\beta_i:i\in[1:j]\}$ for $j\in[2:n]$. Then, $\Omega^{(n)}=\Omega$ and $\Gamma^{(n)}=\Gamma$. Let $F(x)=\det {\rm M}(\Omega,\Gamma\cup\{x\}\setminus\{\beta_n\})$.
Then, $\deg{F(x)}=2n-2$, and the lead coefficient of $F(x)$ is  ${\rm l.c.}(F(x))=\prod^{n-1}_{i=1}\beta_i\cdot\det{\rm M}(\Omega^{(n-1)},\Gamma^{(n-1)}).$

W.L.O.G., we may assume that $0\notin\{\beta_1,\beta_2,\cdots,\beta_{n-1}\}$. According to the definition of $F(x)$,  $\det {\rm M}(\Omega,\Gamma)=F(\beta_n)$ and  $F(\beta_i^{-1})=\beta_i^{-(2n-2)}F(\beta_i)=0$
for $i\in[1:n-1]$, that is, $x-\beta_i \mid F(x)$ and $x\beta_i-1 \mid F(x)$ for $i\in[1:n-1]$. Combining with $\beta^2_i\neq 1$, then $(x-\beta_i)(x\beta_i-1)={\rm lcm}(x-\beta_i,x\beta_i-1)\mid F(x)$. Since $\beta_1,\beta_2,\cdots,\beta_{n-1}$ are distinct in $\mathbb{F}$, we have
${\rm gcd}((x-\beta_1)(x\beta_1-1),\cdots,(x-\beta_n)(x\beta_n-1))=1.$
Then $\prod^{n-1}_{i=1}(x-\beta_i)(x\beta_i-1)|F(x),$ which implies that
\begin{equation}\label{eqh3}
F(x)=\ell(x)\cdot\prod^{n-1}_{i=1}(x-\beta_i)(x\beta_i-1),
\end{equation}
where $\ell(x)\in\mathbb{F}[x]$ is a polynomial. Considering the degree of both sides of equation (\ref{eqh3}), we have $\deg \ell(x)=0$, i.e., $\ell(x) = \ell$ is some non-zero constant in $\mathbb{F}$. Moreover, both sides of equation (\ref{eqh3}) have the same lead coefficient, namely, $$\ell\cdot{\rm l.c.}(\prod^{n-1}_{i=1}(x-\beta_i)(x\beta_i-1))=\prod^{n-1}_{i=1}\beta_i\cdot\det{\rm M}(\Omega^{(n-1)},\Gamma^{(n-1)}),$$
so 
$\ell= {\rm l.c.}(F(x))\cdot\prod^{n-1}_{i=1}\beta^{-1}_i=\det{\rm M}(\Omega^{(n-1)},\Gamma^{(n-1)}).
$
By equation (\ref{eqh3}),
$\det{\rm M}(\Omega,\Gamma)=F(\beta_n)=\det{\rm M}(\Omega^{(n-1)},\Gamma^{(n-1)})\cdot\prod^{n-1}_{i=1}(\beta_n-\beta_i)(\beta_n\beta_i-1).
$
By recursively using such equation, it leads to $\det{\rm M}(\Omega,\Gamma)=\prod_{1\leq i<j\leq n}(\beta_j-\beta_i)(\beta_j\beta_i-1).$
\end{proof}

\section{Proof of Lemma \ref{lem2}}
\label{apd1}
\begin{proof}
Suppose $\Omega^{(j)}=\{x^{j+t}-x^{j-t}:t\in[1:j]\}$ and $\Gamma^{(j)}=\{\beta_i:i\in[1:j]\}$ for $j\in[1:n]$. Then $\Omega^{(n)}=\Omega$ and  $\Gamma^{(n)}=\Gamma$. Let
$G(x)=\det {\rm M}(\Omega,\Gamma\cup\{x\}\setminus\{\beta_n\})$, hence $\deg{G(x)}=2n, {\rm l.c.}(G(x))=\prod^{n-1}_{i=1}\beta_i\cdot\det {\rm M}(\Omega^{(n-1)},\Gamma^{(n-1)}).$

W.L.O.G., we may assume that $0\notin\{\beta_1,\beta_2,\cdots,\beta_{n-1}\}$. According to the definition of $G(x)$, one can easily verify that $G(\beta_i^{-1})=\beta_i^{-(2n-2)}G(\beta_i)=0$
for $1\leq i\leq n-1$. Hence, for $1\leq i\leq n-1 $, $x-\beta_i \mid G(x)$ and $x\beta_i-1 \mid G(x)$. Moreover, when $x^2=1$, the last row of ${\rm M}(\Omega,\Gamma\cup\{x\}\setminus\{\beta_n\})$ is a zero vector, that is, $x^2-1 \mid G(x)$. Since $\beta_i\beta_j\neq 1$ for $1\leq i,j\leq n$ and $\beta_i\neq \beta_j$ for $1\leq i<j\leq n$, $
{\rm gcd}(x^2-1,(x-\beta_1)(x\beta_1-1),\cdots,(x-\beta_n)(x\beta_n-1))=1,
$
which implies that $(x^2-1)\cdot\prod^{n-1}_{i=1}(x-\beta_i)(x\beta_i-)|G(x).$ Hence
\begin{equation}\label{eqH3}
G(x)=\ell(x)\cdot(x^2-1)\cdot\prod^{n-1}_{i=1}(x-\beta_i)(x\beta_i-1),
\end{equation}
where $\ell(x)\in\mathbb{F}[x]$. Considering
the both sides of equation (\ref{eqH3}), it has $\deg \ell(x)=0$ and $\ell(x)\cdot{\rm l.c.}(\prod^{n-1}_{i=1}(x-\beta_i)(x\beta_i-1))
=\prod^{n-1}_{i=1}\beta_i\cdot\det {\rm M}(\Omega^{(n-1)},\Gamma^{(n-1)}),$
that is, $
\ell(x)=\det {\rm M}(\Omega^{(n-1)},\Gamma^{(n-1)}).
$
Then,
$\det {\rm M}(\Omega^{(n)},\Gamma^{(n)})=\det {\rm M}(\Omega^{(n-1)},\Gamma^{(n-1)})
\cdot(\beta_n^2-1)\cdot\prod^{n-1}_{i=1}(\beta_n-\beta_i)(\beta_n\beta_i-1).
$
By recursively using such equation, we get $\det {\rm M}(\Omega,\Gamma)=\prod_{i=1}^n(\beta^2_i-1)\cdot\prod_{1\leq i<j\leq n}(\beta_j-\beta_i)(\beta_j\beta_i-1).$
\end{proof}


\begin{thebibliography}{100}

\bibitem{Alon99}
N. Alon, ``Combinatorial Nullstellensatz," {\it Combinatorics, Probability and Computing}, vol.8,  pp. 7--29, 1999.


\bibitem{APS17}
M. F. Aktas, P. Peng, and E. Soljanin, ``Effective straggler mitigation:
Which clones should attack and when?" {\it ACM SIGMETRICS Perform.
Eval. Rev.}, vol. 45, no. 2, pp. 12--14, 2017.

\bibitem{AAH19}
 N. Azizan-Ruhi, F. Lahouti, A. S. Avestimehr, and B. Hassibi, ``Distributed solution of large-scale linear systems via accelerated projectionbased consensus," {\it IEEE Transactions on Signal Processing}, vol. 67, no. 14, pp. 3806--3817,  2019.

\bibitem{DA22}
A. B. Das and A. Ramamoorthy, ``A Unified Treatment of Partial Stragglers and Sparse Matrices in Coded Matrix Computation," {\it IEEE Journal on Selected Areas in Information Theory}, vol. 3, no. 2, pp. 241-256,  2022.

\bibitem{DAIT22}
 A. B. Das and A. Ramamoorthy, ``Coded sparse matrix computation
schemes that leverage partial stragglers," {\it IEEE Transactions on Information Theory}, vol. 68, no. 6, pp. 4156--4181, 2022.


\bibitem{DAV21}
A. B. Das, A. Ramamoorthy, and N. Vaswani, ``Efficient and robust distributed matrix computations via convolutional coding," {\it IEEE Transactions on Information Theory}, vol. 67, no. 9, pp. 6266--6282, 2021.


\bibitem{DB13}
J. Dean and L. A. Barroso, ``The tail at scale," {\it Communications of the ACM}, vol. 56, no. 2, pp. 74--80, 2013.

\bibitem{DG04}
J. Dean and S. Ghemawat, ``MapReduce: simplified data processing on large clusters,"{\it Sixth USENIX Symposium on Operating System Design and Implementation}, Dec. 2004.

\bibitem{DCG18}
S. Dutta, Z. Bai, H. Jeong, T.M. Low, and P. Grover, ``A unified coded deep neural network training strategy based on generalized polydot codes for matrix multiplication." in {\it Proceedings of IEEE International Symposium on Information Theory(ISIT)}, 2018, pp. 1585--1589.

\bibitem{DCG16}
S. Dutta, V. Cadambe, and P. Grover, ``Short-dot: computing large linear transforms distributedly using coded short dot
products,"  in {\it Proceeding of the 29th International Conference on Neural Information Processing Systems (NIPS)}, Barcelona, Spain, Dec. 2016, pp. 2100--2108.

\bibitem{DCG19}
S. Dutta, V. Cadambe, and P. Grover, ``Short-dot: Computing large linear
transforms distributedly using coded short dot products," {\it IEEE Transactions on Information Theory}, vol. 65, no. 10, pp. 6171--6193, 2019.

\bibitem{MatDotIT20}
S. Dutta, M. Fahim, F. Haddadpour, H. Jeong, V. Cadambe, and
P. Grover, ``On the optimal recovery threshold of coded matrix multiplication,"
{\it IEEE Transactions on Information Theory}, vol. 66, no. 1,
pp. 278--301, 2019.


\bibitem{FCIT21}
M. Fahim and V. R. Cadambe, ``Numerically Stable Polynomially Coded Computing," {\it IEEE Transactions on Information Theory}, vol. 67, no. 5, pp. 2758--2785, 2021.
\bibitem{HA84}
 K.-H. Huang and J. A. Abraham, ``Algorithm-based fault tolerance for
matrix operations," {\it IEEE Transactions on Computer}, vol. 33, no. 6, pp. 518--528, 1984.

\bibitem{softerfpc} Folded Polynomial Codes [Online].
Available: https://github.com/ShineZ9/FoldedPolynomialCodes

\bibitem{HR15}
T. Herault and Y. Robert, Fault-Tolerance Techniques for High Performance Computing. Berlin, Germany: Springer, 2015.

\bibitem{JDCP21}
H. Jeong, A. Devulapalli, V. R. Cadambe and F. P. Calmon, ``$\varepsilon$-Approximate Coded Matrix Multiplication Is Nearly Twice as Efficient as Exact Multiplication," {\it IEEE Journal on Selected Areas in Information Theory}, vol. 2, no. 3, pp. 845--854, 2021.


\bibitem{JJ21}
Z. Jia and S. A. Jafar, ``Cross subspace alignment codes for coded distributed batch computation," {\it IEEE Transactions on Information Theory}, vol. 67, no. 5, pp. 2821--2846, 2021.




\bibitem{LLPPRIT18}
K. Lee, M. Lam, R. Pedarsani, D. Papailiopoulos, and K. Ramchandran, ``Speeding up distributed machine learning using  codes," {\it IEEE Transactions on Information Theory}, vol. 64, no. 3, pp. 1514--1529, 2018.


\bibitem{LSR17}
K. Lee, C. Suh, and K. Ramchandran, ``High-dimensional coded matrix multiplication," in {\it Proceedings of IEEE International Symposium on Information Theory(ISIT)}, Aachen, Germany, Jun. 2017, pp. 2418--2422.


\bibitem{LMA16}
S. Li, M. A. Maddah-Ali, and A. S. Avestimehr, ``Coded distributed
computing: Straggling servers and multistage dataflows,"  in {\it Proceedings of 54th Annual Allerton Conference on Communication, Control, and Computing (Allerton)}, Sep. 2016, pp. 164--171.

\bibitem{RT22}
A. Ramamoorthy and L. Tang, ``Numerically stable coded matrix computations via circulant and rotation matrix embeddings,"{\it IEEE Transactions on Information Theory}, vol. 68, no. 4, pp. 2684–2703, 2022.

\bibitem{RDT20}
A. Ramamoorthy, A. B. Das and L. Tang, ``Straggler-Resistant Distributed Matrix Computation via Coding Theory: Removing a Bottleneck in Large-Scale Data Processing,"  {\it IEEE Signal Processing Magazine}, vol. 37, no. 3, pp. 136--145, 2020.


\bibitem{S12}
R. A. Sadek,``SVD based image processing applications: satate of the art, contributions and research challenges", {\it International Journal of Advanced Computer Science and Applications}, vol. 3, no. 7, pp. 26--34, 2012.

\bibitem{SSC}
https://www.hpcplus.net/science-portal-cloud/index


\bibitem{TLDK17}
R. Tandon, Q. Lei, A. G. Dimakis, and N. Karampatziakis, ``Gradient coding: Avoiding stragglers in distributed learning," in {\it  Proceedings of the 34th International Conference on Machine Learning}, pp. 3368--3376, 2017.


\bibitem{TD22}
L. Tauz and L.Dolecek, ``Variable Coded Batch Matrix Multiplication," {\it IEEE Journal on Selected Areas in Information Theory},  vol. 3, no. 2, pp. 306--320, 2022.


\bibitem{WJG15}
D. Wang, G. Joshi, and G. Wornell, ``Using straggler replication to reduce latency in large-scale parallel computing," {\it ACM
SIGMETRICS Performance Evaluation  Review}, vol. 43, no. 3, pp. 7--11, 2015.




\bibitem{WYZWS21}
S. Wang, Q. Yan, J.Zhang, J.Wang and L.Song,``Coded alternating least squares for straggler mitigation in distributed recommendations," in {\it Proceedings of IEEE International Symposium on Information Theory(ISIT)}, 2021, pp. 1058--1063.

\bibitem{YHGR16}
 N. J. Yadwadkar, B. Hariharan, J. E. Gonzalez, and R. Katz,``Multi-task
learning for straggler avoiding predictive job scheduling," {\it The Journal of Machine Learning Research}, vol. 17, no. 1, pp. 3692--3728, 2016.

\bibitem{YAISIT20}
Q. Yu and A. S. Avestimehr, ``Entangled Polynomial Codes for Secure, Private, and Batch Distributed Matrix Multiplication: Breaking the "Cubic" Barrier," in {\it Proceedings of  IEEE International Symposium on Information Theory (ISIT)}, 2020, pp. 245--250.

\bibitem{YLR19}
Q. Yu, S. Li, N. Raviv, S. M. M. Kalan, M. Soltanolkotabi, and S. A. Avestimehr, ``Lagrange coded computing: Optimal design for resiliency, security, and privacy," in {\it Proceedings of  the 22nd International Conference on Artificial Intelligence and Statistics}, 2019, pp. 1215--1225.

\bibitem{YMA17}
Q. Yu, M. Maddah-Ali, and S. Avestimehr, ``Polynomial codes: an optimal
design for high-dimensional coded matrix multiplication," in {\it Proceedings of the 31'st  International Conference on Neural Information Processing Systems(NIPS)}, 2017, pp. 4406--4416.

\bibitem{YMIT20}
Q. Yu, M. A. Maddah-Ali, and A. S. Avestimehr, ``Straggler mitigation
in distributed matrix multiplication: Fundamental limits and optimal
coding," {\it IEEE Transactions on Information Theory}, vol. 66, no. 3, pp. 1920--1933, 2020.



\bibitem{ZCFSS10}
 M. Zaharia, M. Chowdhury, M.J. Franklin, S. Shenker, and I. Stoica, ``Spark: cluster computing with working sets,"{\it in
Proceedings of the 2nd USENIX HotCloud}, vol. 10, p. 10, June 2010.

\bibitem{ZX21}
J. Zhu and X. Tang, ``Secure Batch Matrix Multiplication From Grouping Lagrange Encoding," {\it IEEE Communications Letters}, vol. 25, no. 4, pp. 1119--1123, April 2021.


\bibitem{Modern13}
J.Von Zur Gathen and J. Gerhard, {\it Modern Computer algebra.} Cambridge university press, 2013.
\end{thebibliography}
\end{document}